\documentclass[twocolumn,twocolappendix]{aastex631}
\usepackage{amsmath,amssymb,graphicx}
\usepackage{epsf,verbatim}
\usepackage{hyperref}
\usepackage{comment}
\usepackage{color}
\usepackage{slashed}
\usepackage{subfigure}
\usepackage{comment}
\usepackage{mdwlist, paralist}
\usepackage{rotating}
\usepackage{bm}
\usepackage{multirow}

\begin{document}

\title{The Odd Dark Matter Halos of Isolated Gas-rich Ultradiffuse Galaxies}
\author[0000-0003-1723-8691]{Demao Kong}
\email{demao.kong@tufts.edu}
\affiliation{Department of Physics and Astronomy, Tufts University, Medford, MA 02155, USA}
\affiliation{Department of Physics and Astronomy, University of California, Irvine, CA, 92697, USA}

\author[ 0000-0001-8555-0164 ]{Manoj Kaplinghat}
\email{mkapling@uci.edu}
\affiliation{Department of Physics and Astronomy, University of California, Irvine, CA, 92697, USA}

\author[0000-0002-8421-8597]{Hai-Bo Yu}
\email{haiboyu@ucr.edu}
\affiliation{Department of Physics and Astronomy, University of California, Riverside, CA 92521, USA}

\author[0000-0002-0447-3230]{Filippo Fraternali}
\affiliation{Kapteyn Astronomical Institute, University of Groningen, Landleven 12, 9747 AD, Groningen, The Netherlands}

\author[0000-0001-5175-939X]{Pavel E. Mancera Pi\~{n}a}
\affiliation{Kapteyn Astronomical Institute, University of Groningen, Landleven 12, 9747 AD, Groningen, The Netherlands}
\affiliation{ASTRON, Netherlands Institute for Radio Astronomy, Postbus, 7900 AA, Dwingeloo, The Netherlands}

\begin{abstract}

We analyze circular velocity profiles of seven ultradiffuse galaxies (UDGs) that are isolated and gas-rich.
Assuming that the dark matter halos of these UDGs have a Navarro-Frenk-White (NFW) density profile or a Read density profile (which allows for constant-density cores), the inferred halo concentrations are systematically lower than the cosmological median, even as low as $-0.6$ dex (about $5\sigma$ away) in some cases. Alternatively, similar fits can be obtained with a density profile that scales roughly as $1/r^2$ for radii larger than a few kiloparsecs. Both solutions require the radius where the halo circular velocity peaks ($R_{\rm max}$) to be much larger than the median expectation. Surprisingly, we find an overabundance of such large-$R_{\rm max}$ halos in the IllustrisTNG dark-matter-only simulations compared to the Gaussian expectation. These halos form late and have higher spins compared to median halos of similar masses. The inner densities of the most extreme among these late-forming halos are higher than their NFW counterparts, leading to a $\sim 1/r^2$ density profile. However, the two well-resolved UDGs in our sample strongly prefer lower dark matter densities in the center than the simulated ones. Comparing to IllustrisTNG hydrodynamical simulations, we also find a tension in getting both low enough circular velocities and high enough halo mass to accommodate the measurements. Our results indicate that the gas-rich UDGs present a significant challenge for galaxy formation models.

 \end{abstract}

\keywords{Low surface brightness galaxies (940), Dark matter (353)}

\section{Introduction}
\label{intro}

Ultradiffuse galaxies (UDGs) are a class of galaxies that have an extremely low luminosity. Compared to usual low surface brightness galaxies, UDGs are more extended in the light distribution and hence fainter. Recent observations have revealed a large number of UDGs in galaxy clusters~\citep{2015ApJ...798L..45V, 2015ApJ...807L...2K, 2016ApJS..225...11Y, 2017ApJ...839L..17J, 2017MNRAS.470.1512W, 2018MNRAS.481.4381M, 2019ApJS..240....1Z, 2020A&A...642A..48I, 2020ApJ...894...75L}; see also~\cite{1984AJ.....89...64B,1988ApJ...330..634I} for earlier studies and~\cite{2018RNAAS...2...43C} for related discussions. UDGs have also been found in the in groups or in the field~\citep{2016AJ....151...96M, 2017ApJ...850..109B, 2017ApJ...842..133L, 2017MNRAS.468.4039R,  2017A&A...607A..79V, 2019ApJ...880...30H, 2019MNRAS.490..566J,  2019MNRAS.488.2143P, 2019MNRAS.486..823R, 2020NatAs...4..246G, 2020MNRAS.495.3636M, 2021MNRAS.500.2049P, 2021ApJ...909...20S}.

The existence of these extreme galaxies in both high- and low-density environments has motivated many studies on their formation mechanisms. For UDGs in clusters and groups, environmental effects, such as ram pressure stripping and tidal stripping, likely play an important role~\citep{2015ApJ...798L..45V,2015MNRAS.452..937Y,2018RNAAS...2...43C, 2018MNRAS.480L.106O, 2019MNRAS.485..382C, 2019MNRAS.487.5272J,  2020MNRAS.491.3496C, 2020PhRvL.125k1105Y,2020MNRAS.497.2786T, 2020MNRAS.494.1848S, 2021NatAs...5.1255B,2022NatAs...6..496M}. For UDGs in the field, they could form in dwarf halos on the distribution tail of high angular momentum~\citep{2016MNRAS.459L..51A, 2017MNRAS.468.4039R, 2019MNRAS.490.5182L,2021MNRAS.502.5370W}. In addition, gas outflows driven by baryonic feedback could change the gravitational potential and expel stars to more external orbits, producing extended and faint galaxies~\citep{10.1093/mnrasl/slw210, 2018MNRAS.478..906C}. It is possible that UDGs are produced owing to a combination of multiple formation channels. For example, both feedback and environmental effects could be important for UDGs in high-density environments~\citep{2019MNRAS.485..796M, 2021MNRAS.502.4262J}.

Kinematic measurements are essential for understanding mass distributions in UDGs and further testing their formation mechanisms, although they are often difficult owning to the low surface brightness nature~\citep{ 2018ApJ...866..112G, 2019MNRAS.484.3425M, 2020MNRAS.495.2582G}. Most isolated UDGs in the field are gas-rich and recent measurements from their neutral gas \textsc{Hi} emissions indicate that their circular velocities are low~\citep{2017ApJ...842..133L, 2020MNRAS.495.3636M}. In fact,~\cite{2021ApJ...909...20S} and ~\cite{2022MNRAS.512.3230M} showed, using high-resolution data, that the host halos of two gas-rich UDGs have low concentrations.

In this work, we focus on a sample of seven gas-rich UDGs in the field: five of them have low-resolution kinematic data from~\cite{2020MNRAS.495.3636M}, and two have high-resolution ones from ~\cite{2022MNRAS.512.3230M, 2021ApJ...909...20S}. We propose a unified model to describe the surface mass density of \textsc{Hi} gas for both high- and low-resolution UDGs, fit their circular velocity data, and infer properties of dark matter halos. We will show that the host halos are ``odd" in the sense that they are extremely low concentration or that their profiles depart dramatically from the Navarro–Frenk–White (NFW) profile in the inner and outer regions or both. We search the IllustrisTNG simulations~\citep{2019ComAC...6....2N} and find an overabundance of such halos, much higher than expected from a Gaussian tail. We will further discuss characteristic properties of the simulated halos and challenges in fully accommodating the measurements of the gas-rich UDGs.

The rest of the paper is organized as follows. In Sec.~\ref{sec:modelling}, we discuss mass modeling. We provide fits to the circular velocity profiles in Sec.~\ref{sec:fits}, and we infer halo parameters and compare them to the IllustrisTNG simulations in Sec~\ref{sec:c200}. We study inner dark matter densities of the halos in Sec~\ref{sec:inner}. We discuss properties of the IllustrisTNG halos in Sec.~\ref{sec:late}. We highlight tensions between the field UDGs and their IllustrisTNG analogs and discuss possible solutions in Sec.~\ref{sec:discussion}. We conclude in Sec.~\ref{sec:conclusions}. In the Appendix, we provide additional information on the mass modeling, fits, and density profiles of the simulated halos.

\section{Mass Modeling}

\label{sec:modelling}

\begin{figure} [!t]
\includegraphics[scale=0.5]{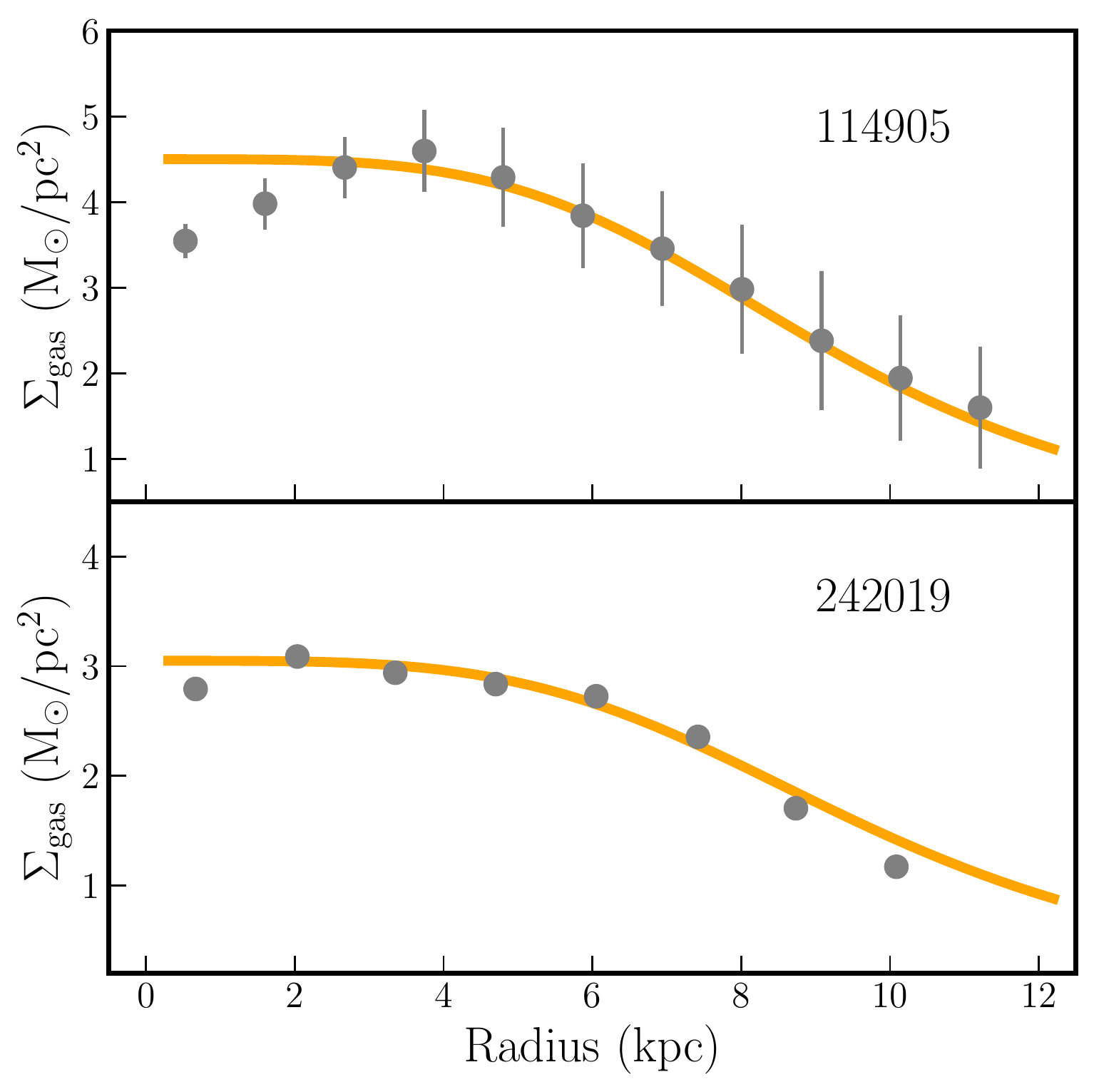}
\caption{Gas surface mass density profiles (solid orange curves) for resolved AGC 114905 (top panel) and AGC 242019 (bottom panel), reconstructed using the gas model in equation~\ref{eq:gas}. The measured profiles are shown for comparison (gray circles), taken from ~\cite{2022MNRAS.512.3230M} and ~\cite{2021ApJ...909...20S}. The helium correction is included.}
\label{fig:gas}
\end{figure}

We consider a sample of seven gas-rich isolated UDGs with measured circular velocity profiles based on their \textsc{Hi} kinematics. Two of them have high-resolution data and their gas surface mass density profiles that are well resolved: AGC 114905~\citep{2022MNRAS.512.3230M} and AGC 242019~\citep{2021ApJ...909...20S}. Five are low resolution (two resolution elements):  AGC 122966, AGC 219533, AGC 248945, AGC 334315 and AGC 749290~\citep{2019ApJ...883L..33M,2020MNRAS.495.3636M}. For these low-resolution UDGs, we propose the following model to describe their \textsc{Hi} gas surface mass density:
\begin{eqnarray}
\label{eq:gas}
\Sigma_{\rm \footnotesize \textsc{Hi}}(R)={\Sigma_{\rm \footnotesize \textsc{Hi}}(0)}{\left[1+(b R/R_{\rm \footnotesize \textsc{Hi}})^m\right]^{-n/m}}
\end{eqnarray}
where $\Sigma_{\rm \footnotesize \textsc{Hi}}(0)$ is the central surface density, $R_{\rm \footnotesize \textsc{Hi}}$ is the radius where $\Sigma_{\rm \footnotesize \textsc{Hi}}(R_{\rm \footnotesize \textsc{Hi}})=1~{\rm M_\odot/pc^2}$, the parameter $b$ is given by the condition $b=\{[\Sigma_{\rm \footnotesize \textsc{Hi}}(0)/{{\rm (M_\odot/pc^2)}}]^{m/n}-1\}^{1/m}$, and we determine numerical factors $n$ and $m$ by comparing the model with the high-resolution UDGs.

For $m=n=4$, we find that the model can reproduce the observed \textsc{Hi} surface mass density of the two UDGs with high-resolution data. In this case, the total enclosed \textsc{Hi} gas mass within the radius $R_{\rm out}$ can be calculated as
\begin{eqnarray}
\label{eq:gasmass}
M_{\rm \footnotesize \textsc{Hi}} (R_{\rm out})=\pi\Sigma_{\rm \footnotesize \textsc{Hi}}(0) \left[\frac{R_{\rm \footnotesize \textsc{Hi}}} {b}\right]^2{\rm Arctan}\left[\left(\frac{bR_{\rm out}}{R_{\rm \footnotesize \textsc{Hi}}}\right)^2\right].
\end{eqnarray}
Fig.~\ref{fig:gas} shows our gas surface mass density profiles for AGC 114905 and AGC 242019 (solid orange curves), compared to measured ones (gray circles), after including helium correction $\Sigma_{\rm gas}=1.33\Sigma_{\rm \footnotesize \textsc{Hi}}$. Our model well reproduces the observations of the two galaxies. For AGC 114905, we fix $R_{\rm \footnotesize \textsc{Hi}}=11.4~{\rm kpc}$, which is $1\sigma$ higher than the median reported in~\cite{2021ApJ...909...19G}. For AGC 242019, $R_{\rm \textsc{Hi}}=10.4~{\rm kpc}$. We have also checked that the enclosed \textsc{Hi} mass calculated using Eq.~\ref{eq:gasmass} agrees well with the measured one for the two UDGs. Our model slightly overestimates the gas surface mass density in the inner region $R\lesssim2~{\rm kpc}$ for AGC 114905. Since this region is dominated by the stellar mass, the discrepancy in $\Sigma_{\rm gas}$ has a negligible effect on the fit.

We apply the model ($m=n=4$) to the five low-resolution UDGs in our sample. We take two approaches to determine $R_{\rm \footnotesize \textsc{Hi}}$ in our fits: fix $R_{\rm \footnotesize \textsc{Hi}}$ to be $1\sigma$ higher than the median for individual UDGs as listed in Table 3 of~\cite{2021ApJ...909...19G}, see our Table~\ref{table:parameters}, motivated by the comparison shown in Fig.~\ref{fig:gas}, and treat $R_{\rm \footnotesize \textsc{Hi}}$ as a varying parameter. We find that the inferred halo properties are almost the same in both approaches. In this paper, we will report the results from the former. We obtain $\Sigma_{\rm gas}(0)$ using the condition in Eq.~\ref{eq:gasmass}, where we fix $R_{\rm out}$ for each of the low-resolution UDGs as listed in Table~\ref{table:parameters}, while varying $M_{\rm \footnotesize \textsc{Hi}}$ (see Sec.~\ref{sec:fits} for details). Fig.~\ref{fig:gas2} in the Appendix shows detailed $\Sigma_{\rm \footnotesize \textsc{Hi}}(R)$ profiles for the low-resolution UDGs after taking representative mode parameters, where we also include profiles with $m=4$ and $n=2.5$ for comparison.

For the stellar component, we use a thin-disk model as~\citep{1970ApJ...160..811F}
\begin{equation}
\label{eq:disc}
\Sigma_{\star}(R)=\Sigma_{\star}(0)\mathrm{e}^{-R / R_{\star}},
\end{equation}
where $\Sigma_{\star}(0)$ is the central surface density and $R_{\star}$ is the scale length of the stellar disk. The total stellar mass can be calculated as $M_{\star}=2\pi\Sigma_{\star}(0)R^2_{\star}$. For the five low-resolution UDGs, we fix $M_\star$ and $R_\star$ to their corresponding median values; see Table~\ref{table:parameters}. For the high-resolution ones,~\cite{2022MNRAS.512.3230M} and ~\cite{2021ApJ...909...20S} have obtained profiles of baryonic circular velocities based on the measured stellar and gas distributions, and we directly use their profiles in our fits.

For the halo, we first use a restricted form of the Read profile~\citep{2016MNRAS.459.2573R}. The enclosed mass within a radius $r$ is given by
\begin{equation}
\label{eq:Read}
M(r)=4\pi\rho_s r^3_s\left[\ln \left(1+\frac{r}{r_s}\right)-\frac{r}{r+r_s}\right] f(r)^\delta,
\end{equation}
where $r_s$ is the scale radius, $\rho_s$ is the scale density, $\delta$ is a numerical factor, and $f(r)=\tanh(r/r_c)$ with $r_c$ being the core radius. For $r\gg r_c$, the function $f(r)$ tends to $1$ and we recover the NFW profile with the same $r_s$ and $\rho_s$.~\citet{2022MNRAS.512.3230M} analyzed AGC 114905 with $\delta$ and $r_c$ being varied according to the relations found in simulations~\cite{2016MNRAS.459.2573R}, and found that $\delta$ is effectively $0$ if the concentration is low. Since those relations are not calibrated for low-concentration halos and current measurements of circular velocities are unlikely to be sensitive to both $r_c$ and $\delta$, we relax the constraints and set $\delta=1$ to ensure a complete transition to a cored profile, while leaving the core radius $r_c$ as a free parameter. This choice allows the fits to explore a broad range of profiles from NFW-like to large core sizes of order $r_s$.

For completeness, we perform additional fits with the Read profile in two extreme limits. By setting $\delta=0$, we consider an exact NFW profile. Moreover, we will also treat both $\delta$ and $r_c$ as free parameters, i.e., a general Read profile. The goal is to demonstrate that the low halo concentration inferred from the gas-rich UDGs is driven by the high \textsc{Hi} gas mass, and it is robust to the variation of the inner density profile. The results of these additional fits will be shown in the Appendix.

One of the benefits of using the Read profile is that we can directly relate $r_s$ and $\rho_s$ to constraints from the cosmological concentration-mass $c_{200}\textup{--}M_{200}$ relation (see~\citealt{2019ApJ...871..168D}), which is often based on fitting simulated halos with an NFW profile. In this work, we will also use the maximal circular velocity $V_{\rm max}$ and its corresponding radius $R_{\rm max}$ to specify a halo. For an NFW-like profile like the Read one, $c_{200}=r_{200}/r_s$ with $r_{200}$ being the virial radius of the halo; $V_{\rm max}=1.64r_s\sqrt{G\rho_s}$, where $G$ is Newton's constant; and and $R_{\rm max}=2.16r_s$.

The assumption of an NFW halo profile in the outer regions may not be a good approximation for halos far below the median concentration. Hence, we also consider a more general double-power-law (DPL) density profile, defined as
\begin{equation}
\label{eq:dpl}
\rho(r) = \rho_s \left(\frac{r}{r_s}\right)^{-\gamma} \left(1+\frac{r}{r_s}\right)^{\gamma-\beta},
\end{equation}
where $\gamma$ and $\beta$ are numerical factors. For $\gamma=1$ and $\beta=3$, we recover the NFW profile. For $\beta\sim2$, $\rho(r)\propto1/r^2$ as $r>r_s$ and the resulting profile of halo circular velocities is nearly flat. We further generalize the calculation of the concentration parameter to $c_{200}=r_{200}/r_{-2}$, where $r_{-2}=(2-\gamma)r_s/(-2+\beta)$ is the radius at which the logarithmic slope of the density profile is $-2$. For the NFW profile with $\gamma=1$ and $\beta=3$, we have the well-known relations $r_{-2}=r_s$ and $c_{200}=r_{200}/r_s$. However, for $\beta\sim2$, $r_{-2}$ can be much larger than $r_s$.

As we will discuss later, for many individual ``low-concentration" halos in the IllustrisTNG simulations that could potentially host the UDGs, $R_{\rm max}$ is close to $r_{200}$ and their profiles of circular velocities are surprisingly flat. Thus, the DPL profile with $\beta\sim2$ is particularly interesting, and we will explicitly show that it provides a good fit to the IllustrisTNG halos in Sec.~\ref{sec:discussion}. Note that for those simulated halos the density drops sharply around the virial radius $r_{200}$ and hence the halo mass $M_{200}=M(r_{200})$ is well defined. Overall, the Read and DPL ($\beta\sim2$) profiles together allow us to test the robustness of our inferences about UDG halos to variations in both the inner and outer density profiles.

 \begin{table*}

\centering

\begin{tabular}{ c |c |c |c |c |c ||c |c |c |c |c |c }
\hline
{\rm AGC ID} &  $\left[\frac{M_\star}{\rm M_{\odot}}\right]$ & $\frac{R_\star}{\rm kpc}$ & $\left[\frac{M_{\rm \footnotesize \textsc{Hi}}}{\rm M_\odot}\right]$ & $\frac{R_{\rm \footnotesize \textsc{Hi}}}{\rm kpc}$ & $\frac{R_{\rm out}}{\rm kpc}$ & $\left[\frac{\rho_s}{\rm M_\odot kpc^3}\right]$ & $\frac{r_s}{\rm kpc}$ & $\frac{r_c}{\rm kpc}$ & $\left[\frac{M_{\rm \footnotesize \textsc{Hi}}}{\rm M_{\odot}}\right]$ &$\left[\frac{M_{200}}{\rm M_{\odot}}\right]$ & $\frac{r_{200}}{r_s}$   \\
\hline
114905 & $8.12$ & $1.79$ & $8.99_{-0.07}^{+0.07}$  & $11.4$ & $12$  & $5.24_{-0.12}^{+0.25}$ &
$23.82_{-6.13}^{+7.40}$ & $6.74_{-3.22}^{+2.29}$ &   -  & $10.14_{-0.11}^{+0.24}$ & $2.11_{-0.28}^{+0.70}$                  \\
122966 & $7.45$ & $4.15$     & $9.03_{-0.05}^{+0.05}$    & $11.2$ &  $14.3$   & $6.11_{-0.72}^{+0.66}$ & $9.29_{-5.01}^{+15.75}$ & $0.59_{-0.41}^{+1.42}$ & $8.97_{-0.07}^{+0.11}$ & $10.08_{-0.19}^{+0.38}$ & $5.76_{-3.16}^{+6.06}$ \\
219533 & $7.82$ & $2.35$ & $9.24_{-0.06}^{+0.06}$        & $14.2$ &  $12.9$   & $5.97_{-0.71}^{+0.62}$ & $12.78_{-6.89}^{+25.76}$ & $0.57_{-0.40}^{+1.35}$ & $9.14_{-0.06}^{+0.10}$ & $10.32_{-0.23}^{+0.51}$ & $4.83_{-2.63}^{+4.44}$  \\
248945 & $8.35$ & $2.08$ & $8.78_{-0.06}^{+0.06}$        & $7.6$ & $11.4$     & $5.48_{-0.28}^{+0.50}$ & $15.73_{-7.23}^{+9.60}$ & $0.85_{-0.65}^{+3.15}$ & $8.7_{-0.08}^{+0.12}$  & $9.92_{-0.18}^{+0.32}$ & $2.84_{-0.82}^{+2.31}$  \\
334315 & $7.65$ & $3.76$ & $9.22_{-0.05}^{+0.05}$        & $14.4$ & $11.2$    & $5.32_{-0.19}^{+0.33}$ & $22.35_{-7.62}^{+9.38}$ & $0.62_{-0.44}^{+1.90}$ & $9.1_{-0.02}^{+0.05}$ & $10.16_{-0.14}^{+0.26}$ & $2.31_{-0.45}^{+1.04}$  \\
749290 & $8.11$ & $2.38$ & $8.95_{-0.05}^{+0.05}$        & $9.7$ & $11.2$     & $5.40_{-0.22}^{+0.40}$ & $18.15_{-7.09}^{+9.33}$ & $0.91_{-0.71}^{+3.60}$ & $8.85_{-0.04}^{+0.08}$ & $10.00_{-0.17}^{+0.30}$& $2.54_{-0.59}^{+1.50}$   \\
242019 & $8.14$ & - & $8.93_{-0.02}^{+0.02}$          & 10.4 & 15.7              & $5.08_{-0.13}^{+0.21}$ & $47.12_{-15.96}^{+14.77}$ & $0.26_{-0.12}^{+0.27}$ & - & $10.76_{-0.22}^{+0.17}$ & $1.75_{-0.25}^{+0.48}$ \\
242019$^\dagger$ & $8.14$ & - & $8.93_{-0.02}^{+0.02}$          & 10.4 & 15.7              & $5.47_{-0.43}^{+0.61}$ & $24.14_{-14.42}^{+8.98}$ & $3.15_{-1.23}^{+1.67}$ & - & $10.39_{-0.35}^{0.44}$ & $3.18_{-1.46}^{+0.44}$ \\
\hline
\end{tabular}

\caption{Parameters used in mass modeling and those inferred from the fits with the Read density profile in Eq.~\ref{eq:Read}.  Columns from left to right: galaxy AGC ID, stellar mass ($\log_{10}M_\star$), scale radius of stellar disks ($R_\star$), \textsc{Hi} gas mass ($\log_{10}M_{\rm \footnotesize \textsc{Hi}}$), scale radius of \textsc{Hi} surface mass densities ($R_{\rm \footnotesize \textsc{Hi}}$), the farthest radius at which the total \textsc{Hi} mass is measured ($R_{\rm out}$); inferred halo scale density ($\log_{10}\rho_s$), scale radius ($r_s$), \textsc{Hi} gas mass ($\log_{10}M_{\rm \footnotesize \textsc{Hi}}$), halo mass ($\log_{10}M_{200}$) and concentration ($c_{200}=r_{200}/r_s$). For AGC 114905, the measurement data shown in the table are from~\citet{2022MNRAS.512.3230M} and we take their baryonic contribution to the total circular velocity, see the left panel of their Fig. 5 (distance $D=67~{\rm Mpc}$, inclination $i=26^{\rm o}$). For AGC 242019, the data are compiled from~\cite{2021ApJ...909...20S} and we directly fit the circular velocity profile of the dark matter halo listed in their Table 2. For the low-resolution UDGs, the \footnotesize \textsc{Hi} measurement and the stellar data are compiled from~\cite{2019ApJ...883L..33M,2020MNRAS.495.3636M,2021ApJ...909...19G}, and we assume the distance and inclination for individual UDGs as in~\citet{2020MNRAS.495.3636M}. We re-calibrated the stellar mass using the mass-to-light-color relation in~\cite{Du_2020}, yielding slightly smaller $M_\star$ values than those reported in~\citet{2020MNRAS.495.3636M}.
}
\label{table:parameters}

\end{table*}

\begin{table*}[!t]
\centering

\begin{tabular}{c||c|c|c|c|c|c|c|c}
\hline

{\rm AGC ID} & $\left[\frac{\rho_s}{\rm M_\odot kpc^3}\right]$ & $\frac{r_{s}}{\rm kpc}$ &  ${\beta}$ & ${\gamma}$ & $\left[\frac{M_{\rm \footnotesize \textsc{Hi}}}{\rm M_{\odot}}\right]$ & $\left[\frac{M_{200}}{\rm M_{\odot}}\right]$  & $\frac{r_{200}}{r_{s}}$   & $\frac{r_{200}}{r_{-2}}$\\
\hline
114905 & $6.25_{-0.07}^{+0.1}$ & $4.15_{-0.51}^{+0.46}$ & $2.05_{-0.03}^{+0.07}$ & $0.09_{-0.07}^{+0.13}$ & - & $9.98_{-0.06}^{+0.12}$ & $10.84_{-0.63}^{+1.44}$ & $0.28_{-0.21}^{+0.41}$ \\
122966 & $7.34_{-0.71}^{+0.59}$ & $1.34_{-0.69}^{+1.9}$ & $2.17_{-0.08}^{+0.09}$ & $0.85_{-0.53}^{+0.45}$ & $8.98_{-0.07}^{+0.1}$ & $10.12_{-0.16}^{+0.22}$ & $37.59_{-20.67}^{+35.58}$ & $5.11_{-2.85}^{+6.19}$ \\
219533 & $7.29_{-0.69}^{+0.6}$ & $1.61_{-0.84}^{+2.25}$ & $2.17_{-0.08}^{+0.09}$ & $0.85_{-0.53}^{+0.44}$ & $9.15_{-0.06}^{+0.09}$ & $10.28_{-0.16}^{+0.23}$ & $35.66_{-19.29}^{+34.84}$ & $4.83_{-2.61}^{+5.39}$ \\
248945 & $6.85_{-0.47}^{+0.74}$ & $1.94_{-1.16}^{+1.67}$ & $2.15_{-0.07}^{+0.09}$ & $0.59_{-0.42}^{+0.53}$ & $8.69_{-0.06}^{+0.11}$ & $9.86_{-0.14}^{+0.23}$ & $21.54_{-9.24}^{+29.2}$ & $2.59_{-1.43}^{+3.64}$ \\
334315 & $6.63_{-0.31}^{+0.54}$ & $3.01_{-1.49}^{+1.48}$ & $2.14_{-0.06}^{+0.1}$ & $0.5_{-0.34}^{+0.46}$ & $9.1_{-0.02}^{+0.04}$ & $10.09_{-0.09}^{+0.18}$ & $16.6_{-5.13}^{+15.04}$ & $1.79_{-0.87}^{+1.6}$ \\
749290 & $6.68_{-0.34}^{+0.63}$ & $2.52_{-1.39}^{+1.49}$ & $2.14_{-0.07}^{+0.09}$ & $0.49_{-0.35}^{+0.5}$ & $8.85_{-0.04}^{+0.07}$ & $9.93_{-0.12}^{+0.23}$ & $17.56_{-5.95}^{+19.37}$ & $1.95_{-1.0}^{+2.19}$ \\
242019 & $6.44_{-0.19}^{+0.29}$ & $5.24_{-1.83}^{+1.76}$ & $2.11_{-0.08}^{+0.11}$ & $0.5_{-0.22}^{+0.16}$ & - & $10.57_{-0.16}^{+0.18}$ & $13.51_{-2.64}^{+5.7}$ & $1.05_{-0.74}^{+0.91}$ \\
242019$^\dagger$ & $6.51_{-0.2}^{+0.2}$ & $5.73_{-1.75}^{+2.21}$ & $2.13_{-0.09}^{+0.11}$ & $0.24_{-0.16}^{+0.19}$ & - & $10.75_{-0.19}^{+0.22}$ & $14.19_{-2.91}^{+4.26}$ & $1.07_{-0.73}^{+0.74}$ \\
\hline
\end{tabular}

\caption{Model parameters inferred from the fits with the DPL density profile in Eq.~\ref{eq:dpl}, including the halo scale density $\log_{10}\rho_s$, scale radius ($r_s$), outer logarithmic slope of the profile ($\beta$), inner logarithmic slope ($\gamma$), \textsc{Hi} gas mass ($\log_{10}M_{\rm \footnotesize \textsc{Hi}}$), halo mass ($\log_{10}M_{200}$), ratio of $r_{200}$ to $r_s$, and concentration ($c_{200}=r_{200}/r_s$). The parameters for modeling stellar and gas distributions are the same as those in Table~\ref{table:parameters}. The inferences for $\beta$ reflect the right prior on it $2.05 < \beta < 2.3$.
}
\label{table:parameters2}
\end{table*}

\begin{figure*} [!t]
\includegraphics[scale=0.3]{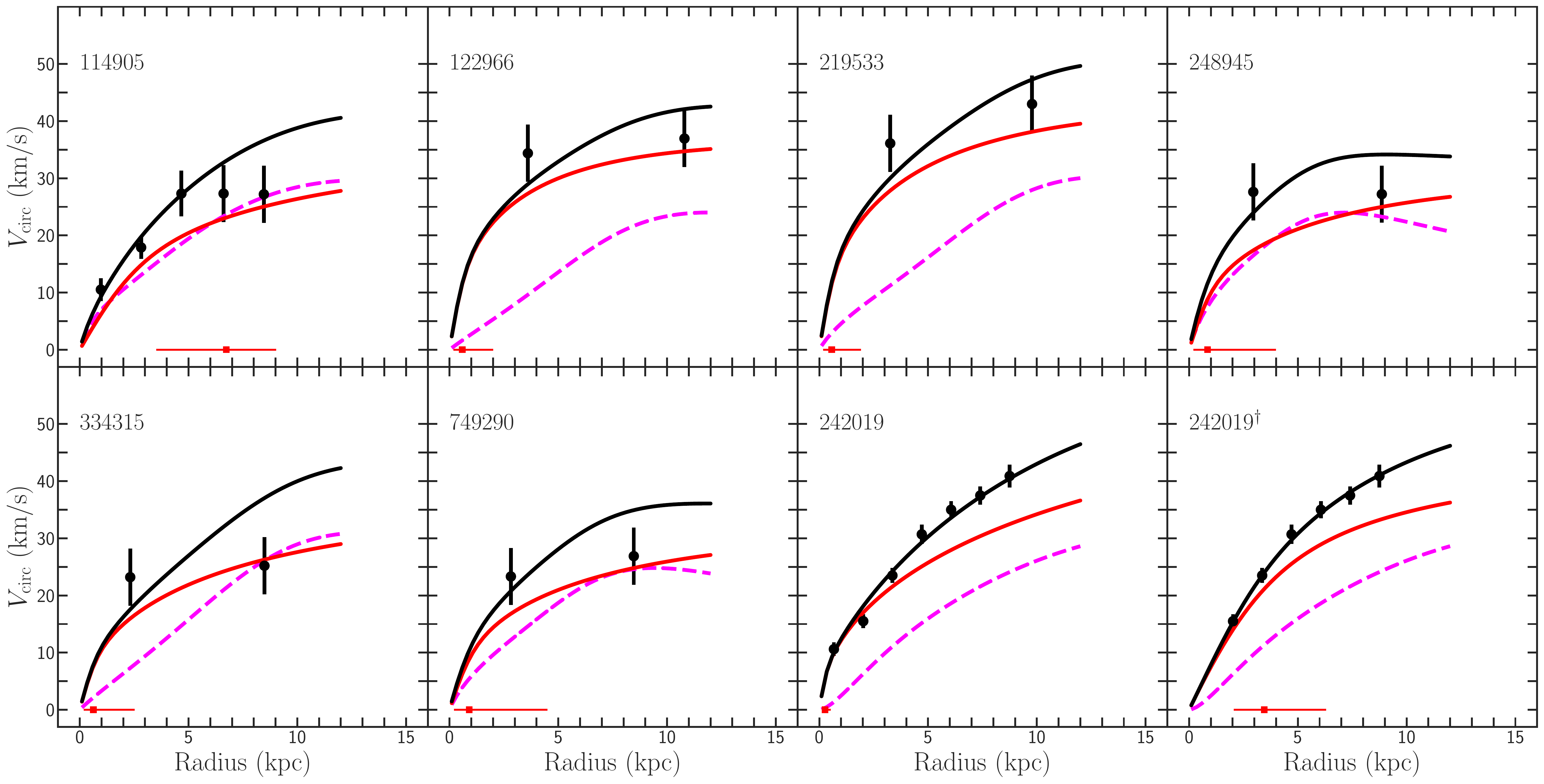}
\caption{Circular velocity profiles of the best fit (solid black), including halo (solid red) and baryonic (dashed magenta) contributions, compared to observational data (black circles), based on the Read profile in Eq.~\ref{eq:Read}. For AGC 242019$^{\dagger}$, the fit is performed without including the innermost datapoint. The horizontal lines denote the halo core size $r_c$ within the $68$th percentile inferred from the fits (red). The observational data are taken from~\cite{2022MNRAS.512.3230M,2020MNRAS.495.3636M} and ~\cite{2021ApJ...909...20S}. For the fits based on the DPL profile in Eq.~\ref{eq:dpl}, see Fig.~\ref{fig:fits2} in the Appendix.}
\label{fig:fits}
\end{figure*}

\section{Fits to the measured circular velocities}
\label{sec:fits}

We calculate the total predicted circular velocity as $V^{\rm mod}_{\rm circ}=\sqrt{V^2_{\rm star}+V^2_{\rm gas}+V^2_{\rm halo}}$, where $V_{\rm star}$, $V_{\rm gas}$ and $V_{\rm halo}$ are contributions from the stars, gas, and dark matter halo, respectively. For AGC 114905, we take the profile of baryonic circular velocities from~\citet{2022MNRAS.512.3230M}; see the left panel of their Fig. 5 ($D=67~{\rm Mpc}$, $i=26^{\rm o}$). For AGC 242019, we directly fit to the halo circular velocities after subtraction of baryonic contributions, as listed in Table 2 of~\cite{2021ApJ...909...20S}. For the low-resolution UDGs, we calculate $V_{\rm gas}$ and $V_{\rm star}$ using the thin-disk approximation with the surface density profiles in Eqs.~\ref{eq:gas} and~\ref{eq:disc}, respectively (\citealt{2008gady.book.....B}).

 For the contribution from gas,
\begin{eqnarray}
V^2_{\rm gas}=-4G\int^R_0 \frac{a da}{\sqrt{R^2-a^2}}\frac{d}{da}\int^\infty_a \frac{R'\Sigma_{\rm gas}(R')dR'}{\sqrt{R'^2-a^2}}.
\label{eq:vgas}
\end{eqnarray}
In practice, we convert Eq.~\ref{eq:vgas} into a dimensionless form and generate numerical templates for the fits. For the stellar component, $V_{\rm star}$ has an analytical form
\begin{eqnarray}
V^2_{\rm star}=4\pi G\Sigma_\star(0)R_\star y^2[I_0(y)K_0(y)-I_1(y)K_1(y)],
\end{eqnarray}
where $y\equiv R/2R_\star$, and $I_{0,1}$ and $K_{0,1}$ are modified Bessel functions.

We use an Markov Chain Monte Carlo (MCMC) routine {\tt emcee}~\citep{2013PASP..125..306F} to fully explore the halo parameters and find the best fit that minimizes the residuals of $V^{\rm mod}_{\rm circ}$. The likelihood function is $\exp(-0.5\chi^2)$, where $\chi^2={(V^{\rm mod}_{\rm circ}-V^{\rm obs}_{\rm circ})^2/\delta V^2_{\rm circ}}$. We impose the following flat priors on the relevant parameters. For the Read profile, we impose $7\leq\log_{10} (M_{200}/{\rm M_{\odot}})\leq12$, $0\leq\log_{10}c_{200}\leq 1.7$, $-1\leq\log_{10}(r_c/{\rm kpc})\leq1$, and $r_c/r_s<1$. For the DPL profile, we have the same prior on $M_{200}$, while for the other parameters $1\leq\log_{10}(r_{200}/r_s)\leq 2$, $0<\gamma<1.5$, and $2.05<\beta<2.3$. Both the $r_s$ and $\beta$ priors are chosen to recover a profile of $\rho\propto1/r^2$ behavior as evident within $10~{\rm kpc}$, where we have the measurement data. The prior on the inner slope $\gamma$ is generous and allows the profile to be cored or cuspy.

For the low-resolution UDGs, we impose a flat prior on $\log (M_{\rm \footnotesize \textsc{Hi}}/{\rm M_\odot})$ within the $\pm3\sigma$ range. In addition, we require that $M_{200}$ is large enough such that the cosmological limit $(M_{\rm gas}+M_{\star})/M_{200}\leq\Omega_{b}/(\Omega_m-\Omega_b)=0.187$ is satisfied~\citep{2020A&A...641A...6P}. This puts a conservative lower limit on the halo mass for a given total baryon mass. We summarize the parameters used for mass modeling and those inferred from our fits in Tables~\ref{table:parameters} and~\ref{table:parameters2}.

Fig.~\ref{fig:fits} shows circular velocity profiles of the total best fit (black), including halo (red) and baryonic (magenta) contributions, based on the Read profile in Eq.~\ref{eq:Read}. Compared to the measured kinematic data (black circles), our constructed model works reasonably well. We notice that except for AGC 242019, the fits overestimate the circular velocity at large radii because of significant contributions from the gas disk. In particular, for AGC 114905, AGC 248945, and AGC 749290, the gas and halo contributions are comparable at the radius where the last datapoint was measured (close to $R_{\rm \footnotesize \textsc{Hi}}$). This reflects the fact that these UDGs have a high baryon content~\citep{2019ApJ...883L..33M}. We note that $M_{200}$ is close to the lower limit set by the cosmological baryon fraction prior because the circular velocity data push the dark matter contribution to be small. The best fits with the DPL profile are similar to the fits with the Read profile (Fig.~\ref{fig:fits2} in Appendix). The inferred halo masses are similar as well, because of the strong influence of the lower limit on $M_{200}/(M_{\rm gas}+M_{\star})$.

In Fig.~\ref{fig:fits}, we also show the core size $r_c$ inferred from the fits (red horizontal line). In all low-resolution cases, core sizes of a few kiloparsecs are allowed. But it cannot be larger because circular velocities between $2$ and $8~{\rm kpc}$ do not change significantly. The core sizes can also be zero and hence consistent with an NFW profile. AGC 114905 allows for a core size of $r_c\sim7~{\rm kpc}$, but the other high-resolution one, AGC 242019, prefers effectively zero core size~\citep{2021ApJ...909...20S}. The preference is largely driven by the innermost datapoint of AGC 242019, which has a small error of $1.2~\rm km s^{-1}$. When we remove this datapoint (denoted by $242019^\dagger$ in the legend), there is mild evidence for a nonzero core size of $r_c\sim4~{\rm kpc}$. Removing the innermost datapoint also brings AGC 242019 in agreement with the other UDGs in terms of the preferred $V_{\rm max}$ and $M_{200}$ values as we discuss below. This highlights the importance of accurate measurements at radii below $5~{\rm kpc}$.

\section{Inferred halo parameters compared to simulations}
\label{sec:c200}

\begin{figure*} [!t]
\includegraphics[scale=0.4]{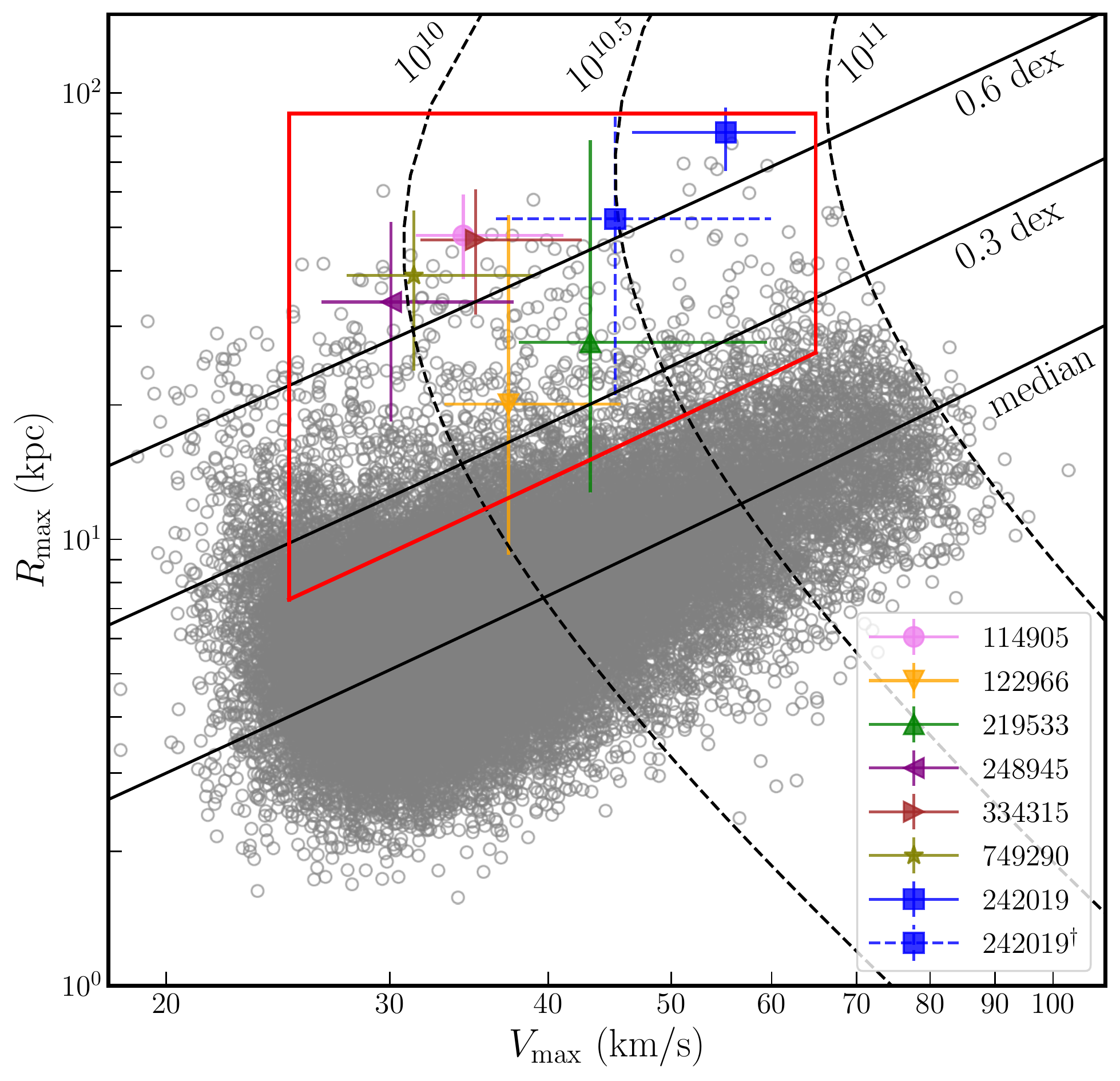} ~~~~~
\includegraphics[scale=0.4]{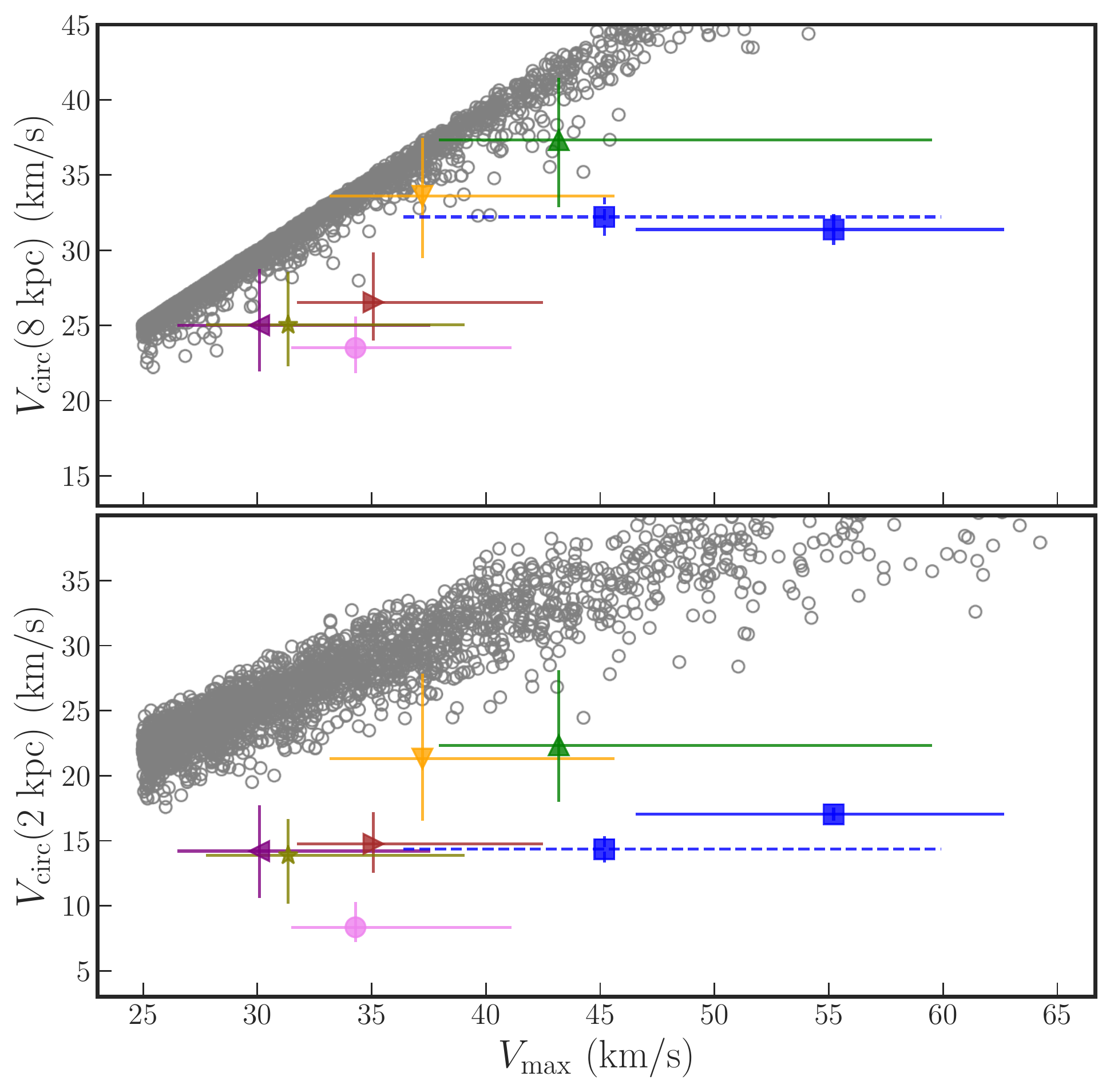}
\caption{{\it Left:} $R_{\rm max}\textup{--}V_{\rm max}$ distributions inferred from the Read fits (colored), together with those from the TNG50-1-Dark simulations (gray circles). The solid black lines show the median relation~\citep{2019ApJ...871..168D}, as well as $0.3$ dex and $0.6$ dex larger $R_{\rm max}$ (``lower concentration") compared to the median. The dashed black lines denote the halo masses $\log_{10}(M_{200}/{\rm M_\odot})=10,~10.5$ and $11$. The IllustrisTNG halos within the red box are selected for further study in the right panel. {\it Right:} the halo circular velocity at radii $8~{\rm kpc}$ (top panel) and $2~{\rm kpc}$ (bottom panel) vs. maximum circular velocity for the observed UDGs (symbols colored according to the legend in the panel on the left), compared to the selected TNG50-1-Dark halos (gray circles).}
\label{fig:rmaxvmax}
\end{figure*}

\begin{figure*}
\includegraphics[scale=0.4]{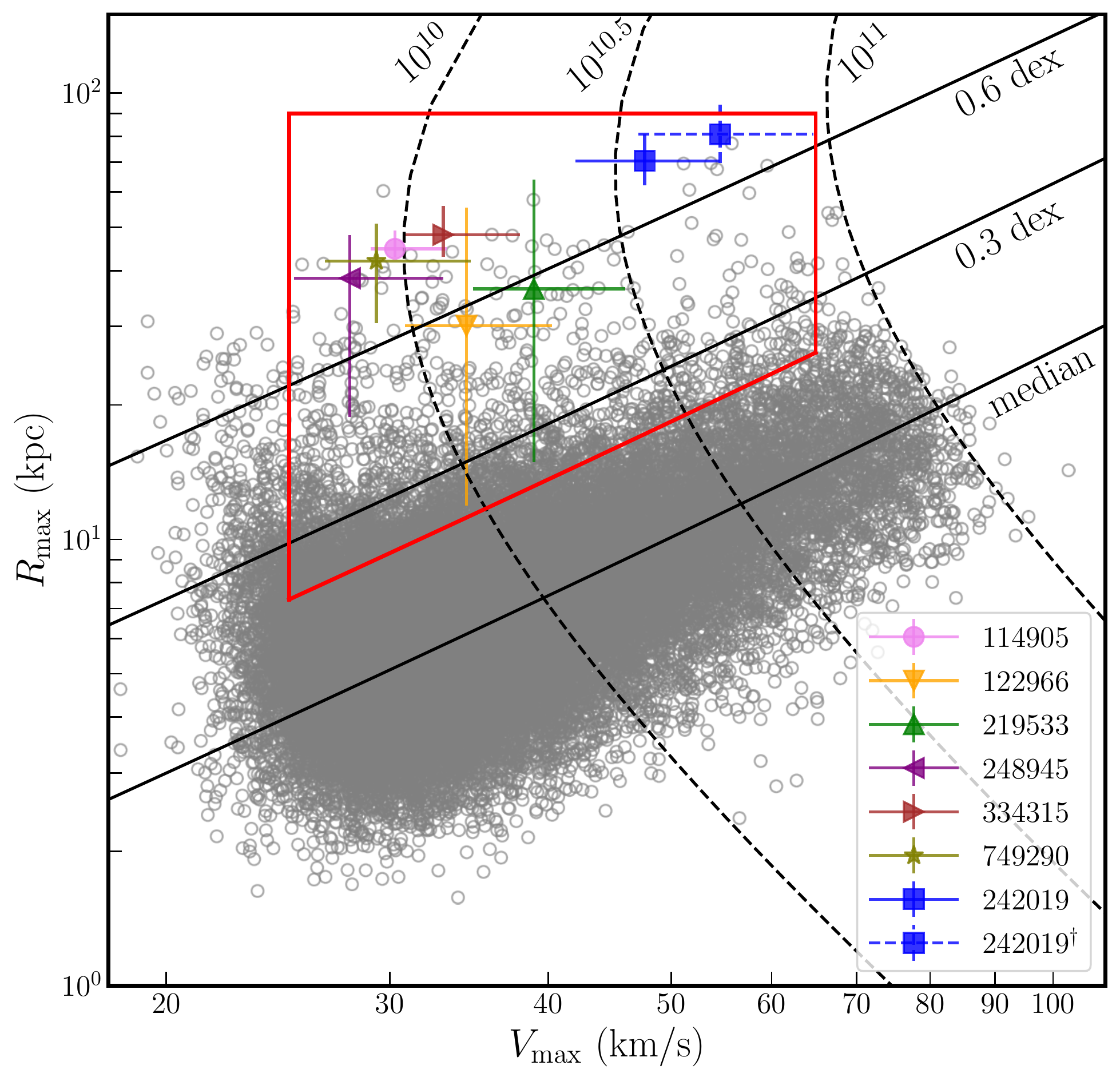} ~~~~~
\includegraphics[scale=0.4]{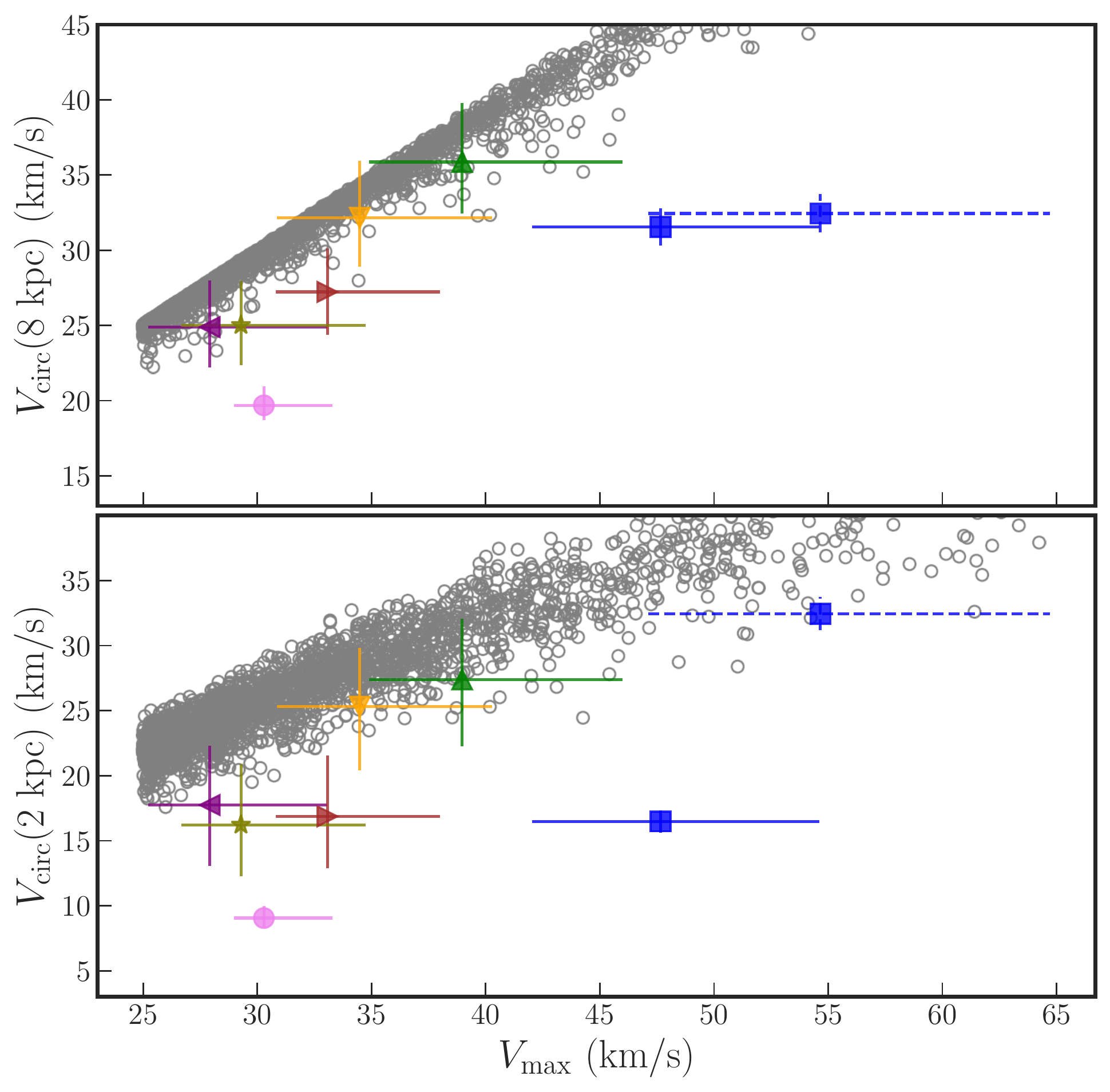}
\caption{{\it Left:} $R_{\rm max}\textup{--}V_{\rm max}$ distributions inferred from the DPL fits (colored), together with those from TNG50-1-Dark simulation (gray circles). The solid and dashed black lines and the red box are the same as in Fig.~\ref{fig:rmaxvmax}. {\it Right:} the halo circular velocity at radii $8~{\rm kpc}$ (top panel) and $2~{\rm kpc}$ (bottom panel) vs. maximum circular velocity for the observed UDGs (symbols colored according to the legend in the panel on the left), compared to the TNG50-1-Dark halos (gray circles) within the red box.}
\label{fig:rmaxvmax2}
\end{figure*}

We will compare the halo parameters inferred from our fits with those from the IllustrisTNG project in~\cite{2019ComAC...6....2N}, which implements a comprehensive model for galaxy formation physics. We take the snapshot $99$ ($z=0$) in the TNG50-1-Dark dataset based on dark-matter-only simulations, which is publicly available, and choose simulated halos with the mass $\log_{10}(M_{200}/{\rm M_{\odot}})=9.3\textup{--}11$ (Primary Flag=1). As we will show later, this mass range is appropriate for the UDGs we consider. In total, there are $37411$ halos, and we retrieve their properties, including concentrations, density profiles, formation histories, and spins.

We choose the TNG50-1-Dark simulated halos for comparison becuase they provide a clean benchmark sample where baryonic effects on the halo properties are absent. For the IllustrisTNG hydrodynamical simulations, supernova feedback is ``weak" in the sense that it does not lead to a lowering of the inner dark matter densities~\citep{2018MNRAS.481.1950L}. Thus, the results from the dark-matter-only simulations should provide a good approximation for the purposes of the current work. We will leave a detailed comparison with the full hydrodynamical TNG50 simulations for future work.

Fig.~\ref{fig:rmaxvmax} (left) shows $R_{\rm max}$ and $V_{\rm max}$ of the halos of the UDGs inferred from the Read fits (colored). If $R_{\rm max}$ is larger than $r_{200}$, we use $r_{200}$ instead. For the UDGs we consider, the values of their halo parameters are $V_{\rm max}\sim30\textup{--}50~{\rm km/s}$, $R_{\rm max}\sim20\textup{--}100~{\rm kpc}$, and $M_{200}\sim10^{10}\textup{--}10^{11}~{\rm M_\odot}$. For comparison, we plot the $R_{\rm max}\textup{--}V_{\rm max}$ median relation predicted in cosmological simulations~\citep{2019ApJ...871..168D}, as well as the lines with $R_{\rm max}$ values $0.3$ dex and $0.6$ dex larger than the median for given $V_{\rm max}$ (solid black). Since the $1\sigma$ scatter for the lognormal $c_{200}\textup{--}M_{200}$ distribution is $0.11$ dex, the concentration of the UDG halos would be $\sim3\sigma\textup{--}5\sigma$ below the cosmological median when if we were to extrapolate the lognormal distribution to the low end. Thus, we conclude that the UDG halos must have extremely low concentrations (large $R_{\rm max}$) if they are hosted in NFW-like halos; see also~\citet{2022MNRAS.512.3230M,2021ApJ...909...20S} for the two high-resolution UDGs. We reach the same conclusion if we adopt the NFW profile or the general Read profile by varying both $r_c$ and $\delta$ simultaneously; see the Appendix.

We have further checked that the conclusion holds for systematically denser gas surface densities, for example, with $n=2.5$ in Eq.~\ref{eq:gas} (see Fig.~\ref{fig:rmaxvmaxn25} in the Appendix), and it holds even when only including the outmost measured datapoint in the fits. The preference for a large $R_{\rm max}$ solution does not depend on whether the profile is cored or cuspy, as we have marginalized over the core radius in the Read profile. We have checked that the core radius is not strongly correlated with other model parameters. From Fig.~\ref{fig:rmaxvmax} (left), we also see that removing the innermost datapoint brings AGC 242019 more in line with the other UDGs in the $R_{\rm max}\textup{--}V_{\rm max}$ plane.

If a Gaussian tail of the $c_{200}\textup{--}M_{200}$ distribution is extended to the low-concentration end, we would not expect to find halos with concentrations as low as $0.5$ dex below the median. However, the Gaussian extrapolation is unwarranted. Fig.~\ref{fig:rmaxvmax} (left) shows results from the TNG50-1-Dark simulations (gray circles), where we have imposed the selection condition $\log_{10}(M_{200}/{\rm M_{\odot}})=9.3\textup{--}11$ corresponding to $V_{\rm max}\sim20\textup{--}80~{\rm km/s}$. There is clearly a non-Gaussian tail to low concentrations that dominates about $0.2$ dex below the median; see also Fig.~\ref{fig:tng} (left). In the $50 h^{-1} \rm Mpc$ box, we find $956$ halos within this mass range with concentrations of $0.3$ dex below the median and $195$ halos with concentrations $0.5$ dex below the median.

The gas-rich UDGs strongly favor halos with anomalously large $R_{\rm max}$, and in many cases inferred $R_{\rm max}$ is close to the virial radius $r_{200}$. This brings up the possibility that the halo density profile could be close to $1/r^2$, which would lead to a flat profile of circular velocities. To investigate this possibility, we have intended to choose the prior for the outer slope in a narrow range of $2.05<\beta<2.3$ for the DPL fits, a significant difference from the NFW one, $\beta=3$. We further take $1\leq \log(r_{200}/r_{s})\leq2$, again with the intention of picking solutions leading to flat halo $V_{\rm circ}$. Fig.~\ref{fig:rmaxvmax2} (left) shows $R_{\rm max}$ and $V_{\rm max}$ of the halos, based on the DPL fits (colored), where $r_{200}$ is reported if $R_{\rm max}>r_{200}$. We see that the inferred $V_{\rm max}\textup{--}R_{\rm max}$ values are consistent with those from the fits with the Read profile.

It is also useful to compare the concentration values inferred from the Read and DPL fits, as summarized in Tables~\ref{table:parameters} and~\ref{table:parameters2}. Both fits show that the halo concentration has to be order unity, and they follow a similar variation pattern for individual UDGs. The $r_s$ values from the DPL fits are much smaller than those from the Read fits. Both fits give rise to similar halo masses owing to the constraint from the cosmological baryon abundance, and the DPL fits require smaller $r_s$ for given $M_{200}$ as the density drops more slower for $r>r_s$, compared to the Read fits.

Based on the fits above, a consistent picture emerges, i.e., the gas-rich UDGs are hosted by halos with $R_{\rm max}$ much larger than the median. This would imply anomalously low concentrations if the halo profile is NFW-like, allowing for a flat density core. On the other hand, equally good fits can be obtained with halo density profiles that are approximately $1/r^2$ and transition to another power law in the inner regions. In both cases, it is clear that the UDG hosts are very different from the median halos in this mass range.

\section{Inferred internal densities compared to simulations}
\label{sec:inner}

To study the large-$R_{\rm max}$, low-concentration IllustrisTNG halos that are overabundant, we selected the simulated halos within the red box of Fig.~\ref{fig:rmaxvmax} (left), which are similar to those hosting the UDGs, and examined their dark matter distributions. The bottom edge of the red box corresponds to a concentration of $0.2$ dex below the median (assuming an NFW profile). Fig.~\ref{fig:rmaxvmax} (right) shows the circular velocities at radii $8$ and $2~{\rm kpc}$ versus $V_{\rm max}$ for the halos inferred from the fits (colored), compared to the simulated ones that are calculated using the mass profiles of the IllustrisTNG halos (gray circles).  The simulated halos shown in Fig.~\ref{fig:rmaxvmax} are all well resolved at $2~\rm kpc$ using the~\citet{2003MNRAS.338...14P} criteria. About $4\%$ of the halos have resolution radii close to $2~{\rm kpc}$, and in all cases the resolution radius is smaller than $2.3~{\rm kpc}$.

We see that while the simulated halos get close to the required densities at $8~\rm kpc$, they are too dense at $2~{\rm kpc}$ compared to the Read fits. The preference for a low density is mainly because the host halos have a low concentration, while a cored profile could also contribute to lowering the central density. All of the UDGs are consistent with core sizes of a few kiloparsecs, except for AGC 242019. For this UDG, the circular velocity at $2~{\rm kpc}$ is about a factor of $2$ smaller than that expected for the simulated halos despite being consistent with zero core size. As discussed in Sec.~\ref{fig:fits}, AGC 242019 allows to have a core with $r_c\sim6~{\rm kpc}$, without including the innermost datapoint at $0.5~{\rm kpc}$. Indeed, AGC 242019$^\dagger$ has lower halo $V_{\rm circ}$ at $2~{\rm kpc}$, and $V_{\rm max}$ becomes smaller as well. Thus, the tension is reduced, but it still remains compared to the simulated halos.

Fig.~\ref{fig:rmaxvmax2} (right) shows the inferences from the DPL fits. We remind the reader that the fits impose a strong prior on the outer logarithmic density slope $2.05<\beta<2.3$, leading to a flat profile of halo circular velocities.
The DPL profile has the freedom to be steeper than the NFW profile in the inner region through the parameter $\gamma$; however, we do not see a systematic shift to $\gamma>1$ in those fits (see Table~\ref{table:parameters2}). The inferred halo circular velocities at $2$ and $8~{\rm kpc}$ are higher than those obtained with the Read fits, resulting in better agreement with the simulated halos. However, overall the simulated halos are still too dense. For the five low-resolution UDGs, their halo circular velocities at $2~{\rm kpc}$ are systematically lower than the simulated ones, although the tension is reduced compared to the Read fits. The high-resolution AGC 242019 and AGC 114905 are still highly discrepant with the TNG50-1-Dark halos. We expect that the five low-resolution UDGs would follow a similar trend if future measurements could resolve the rising part of their inner circular velocity profile.

For both Read and DPL fits, we have imposed a conservative prior $M_{\rm baryon}/M_{200}\leq0.187$, the cosmological baryon fraction. This prior has a large influence on the inferences for the halo parameters. If the halo mass were allowed to be lower, then $V_{\rm max}$ would be lower and more in line with the TNG50-1-Dark simulations shown in the right panels of Figs.~\ref{fig:rmaxvmax} and~\ref{fig:rmaxvmax2}. Thus, the tension between the inferred UDG halos and the simulated ones in the halo $V_{\rm circ}\textup{--}V_{\rm max}$ plane could be related to the prior on the halo mass from the cosmological baryon fraction. In Sec.~\ref{sec:discussion}, we will show examples after relaxing this constraint.

\begin{figure}
\centering
\includegraphics[scale=0.4]{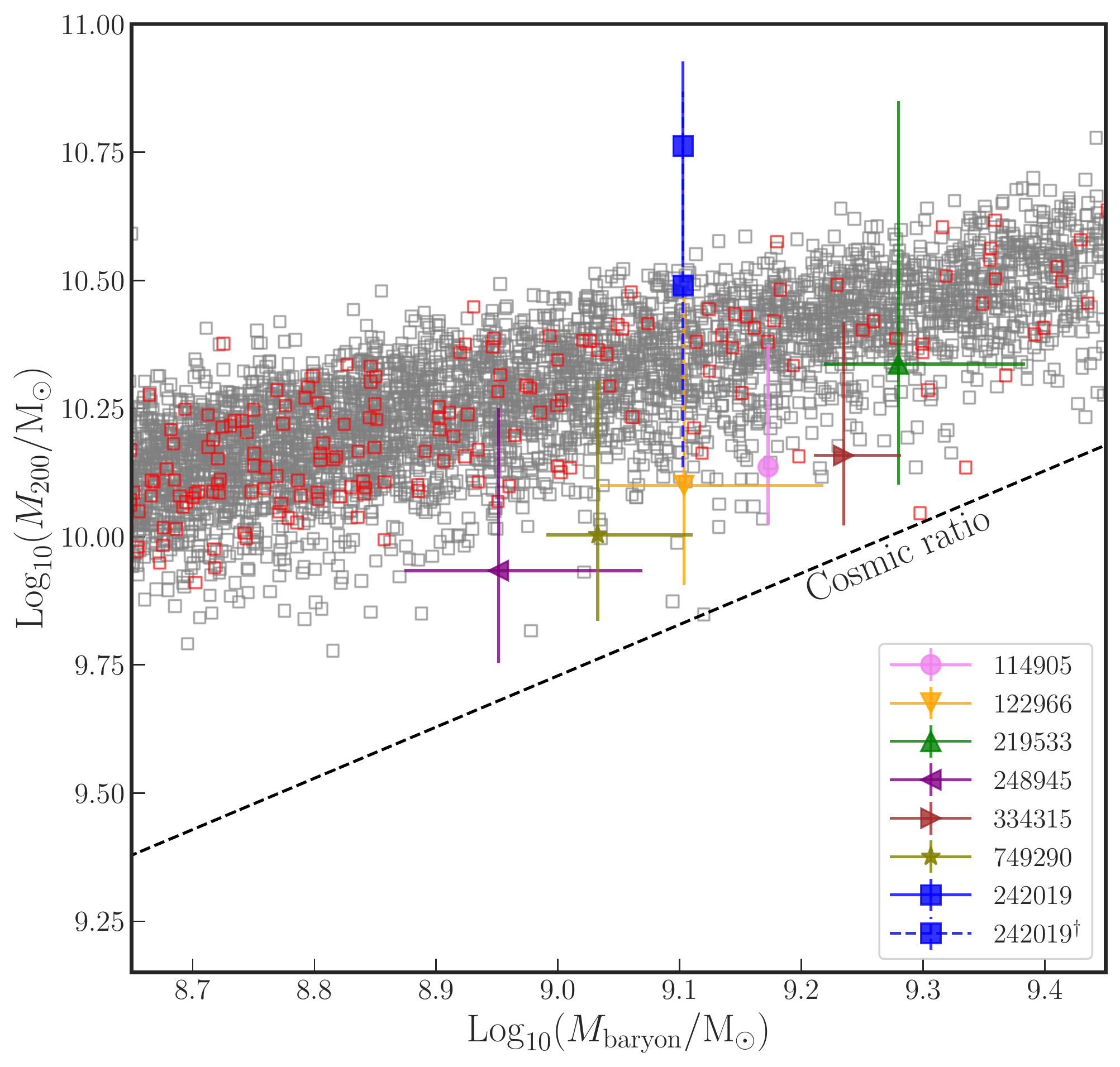}
\caption{Inferred $M_{200}$ vs. $M_{\rm baryon}$ for the UDGs (colored with error bars), compared to selected TNG50-1 galaxies satisfying $M_{\rm gas}/M_{\rm baryon}>0.9$ in a similar mass range (full hydro; gray and red squares). The dashed line denotes the lower limit on the halo mass from the cosmological ratio of baryonic to dark matter masses. The red squares denote low-concentration halos satisfying $\log_{10}(c_{200}/c^{\rm median}_{200})<-0.3$, see Fig.~\ref{fig:tng}. }
\label{fig:m200mgas}
\end{figure}

\begin{figure*} [!t]
\includegraphics[scale=0.3]{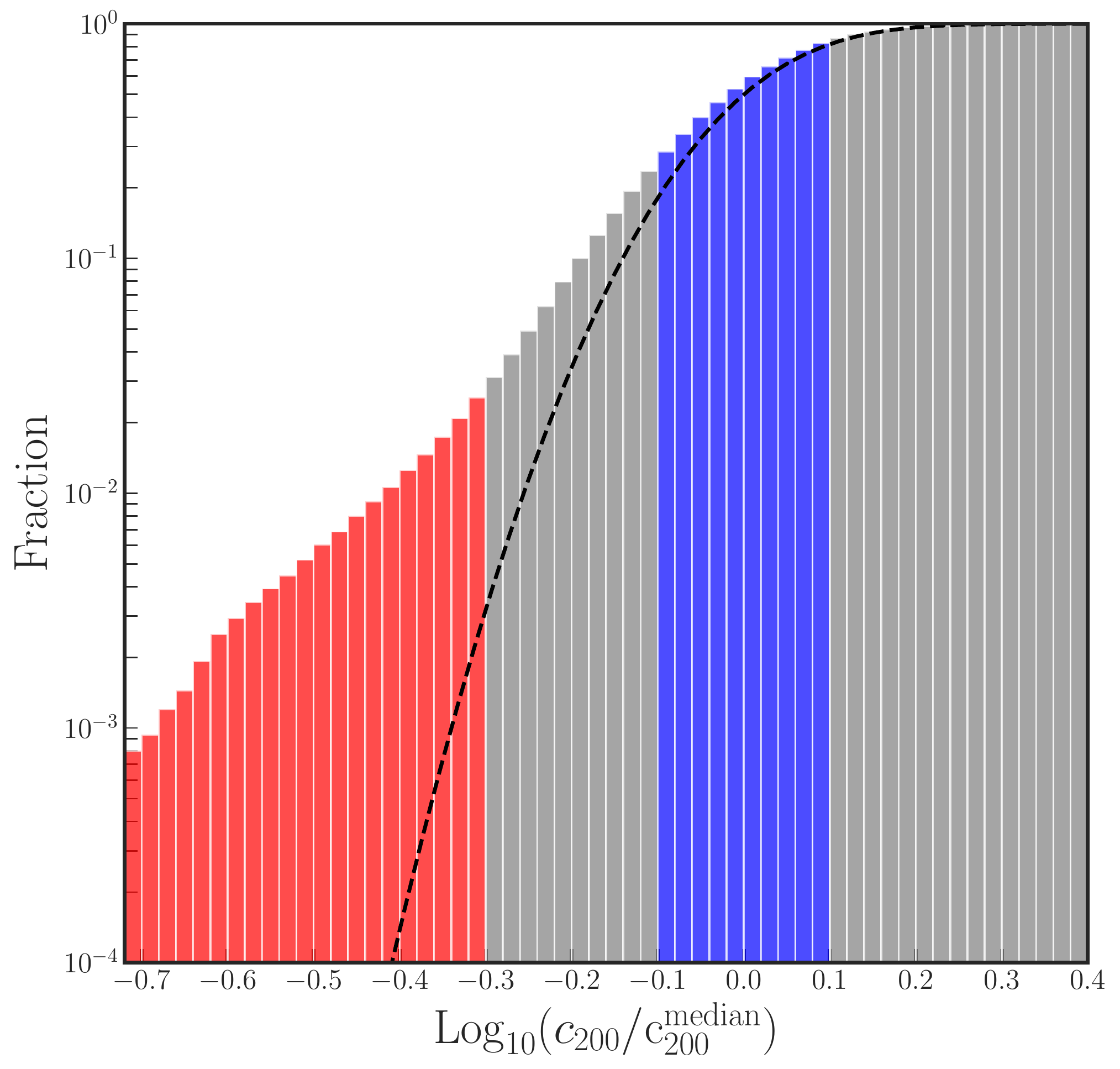}
\includegraphics[scale=0.3]{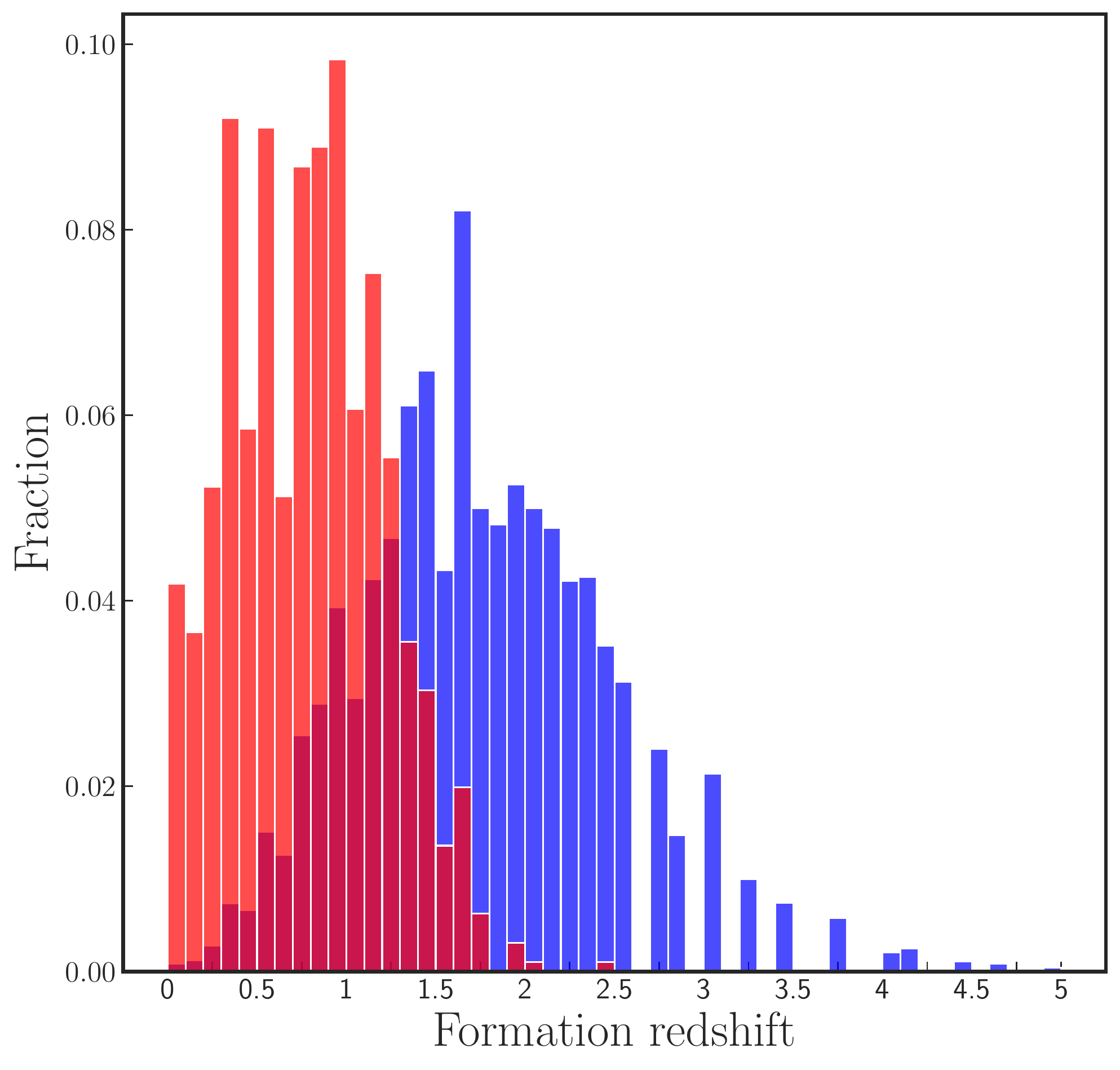}
\includegraphics[scale=0.3]{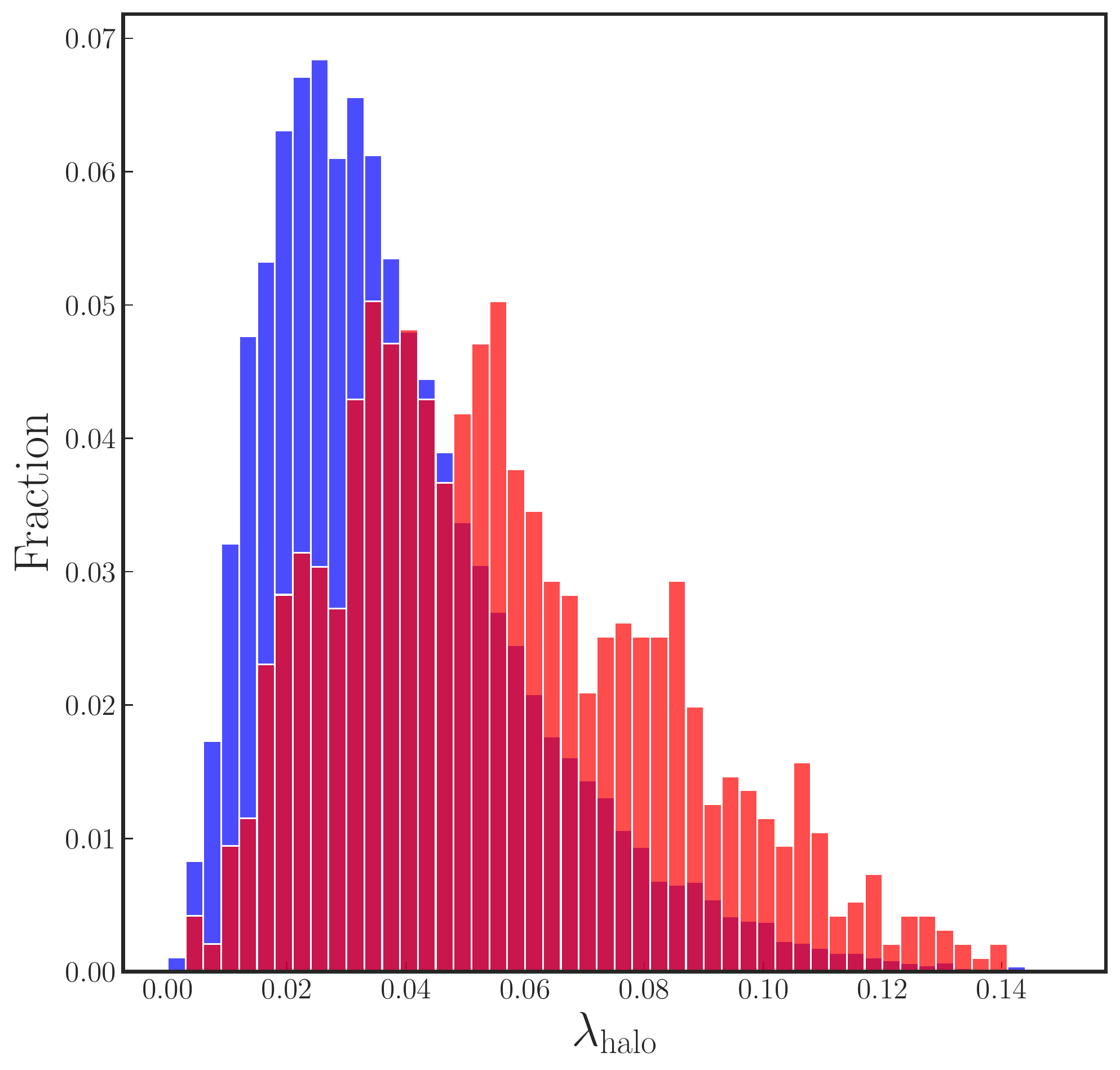}
\caption{{\it Left:} The histogram of ``halo concentration" of the TNG50-1-Dark halos with $\log_{10}(M_{200}/{\rm M_{\odot}})=9.3\textup{--}11$ inferred assuming an NFW profile, compared to the extrapolation with a lognormal distribution~\citep{2019ApJ...871..168D} (dashed black). The red and blue colors highlight two subsets of the halos satisfying $\log_{10}(c_{200}/c^{\rm median}_{200})<-0.3$ and $|\log_{10}(c_{200}/c^{\rm median}_{200})|<0.1$, respectively. Note that many of the systems in the tail have density profiles that fall much slower than $1/r^3$ in the outer regions leading to larger $R_{\rm max}$ and hence lower $c_{200}$. {\it Middle:} the histogram of the redshift at which half of the halo mass is assembled, for the low (red) and median (blue) concentration halos. {\it Right:} the histogram of the total halo spin parameter for the low (red) and median (blue) concentration halos.}
\label{fig:tng}
\end{figure*}

To amplify this point regarding the baryon mass fraction, we check the full hydrodynamic TNG50-1 (highest-resolution TNG50 box) simulations to compare $M_{200}\textup{--}M_{\rm baryon}$ distributions of the simulated galaxies with those from our fits; see Fig.~\ref{fig:m200mgas} (gray and red squares). To focus on extremely gas-rich systems similar to many of the observed field UDGs, where the ratio of total gas-to-stellar masses ranges from a factor of $3$ to $50$, we have selected the simulated galaxies in a similar mass range to the UDGs with a further cut of $M_{\rm gas}/M_{\rm baryon}>0.9$. About $76\%$ of the TNG50-1 galaxies in the queried mass range satisfy the cut, and their baryon masses are similar to those plotted in Fig.~\ref{fig:m200mgas}. In those TNG50-1 galaxies, \textsc{Hi} is the dominant gas component~\citep{2018ApJS..238...33D}.

As shown in Fig.~\ref{fig:m200mgas}, for a given total baryon mass, the inferred $M_{200}$ values of six UDGs are close to the lower end in the $M_{200}\textup{--}M_{\rm baryon}$ plane, and the simulated galaxies have a systematically higher halo mass. Given the uncertainty of the halo mass from the fits, we see that some of the simulated galaxies can be compatible with the observed ones. AGC 242019 favors a high halo mass owing to its sharp rising circular velocities, and again it becomes more aligned with the other UDGs, as well as the simulated ones, if the innermost datapoint is not included. The red squares in Fig.~\ref{fig:m200mgas} denote TNG50-1 halos with the concentration $0.3$ dex below the median. It seems that there is no correlation between halo concentration and gas content. Overall, we find that the IllustrisTNG simulations could produce field UDG analogs in low-concentration halos with high gas content, along with similar halo and gas masses to those of our sample. But we need an additional mechanism to reduce the inner densities to be consistent with the measurements.

Another important factor contributing to the tension between the TNG50-1-Dark simulations and the inferred circular velocities at $2~{\rm kpc}$ (for both DPL and Read fits) is that some of the simulated halos have a steep density profile in the inner regions. We find that this is the case for halos with the largest $R_{\rm max}$, relative to the cosmological median; see Fig.~\ref{fig:halo} in the Appendix for illustrative examples. For these halos, the circular velocity at $2~{\rm kpc}$ is higher than it would be for an NFW profile with the same $V_{\rm max}$ and $R_{\rm max}$. It is likely that the strong deviation from the NFW profile for these outlier halos is due to the fact that they have not yet relaxed. As we look at halos less than $-0.3~\rm dex$ from the median, we find that the halo profiles start to look more NFW-like. We will further discuss these points in the next section.

In short, the internal densities of the UDGs measured at $2~{\rm kpc}$ and $8~{\rm kpc}$ are inconsistent with the TNG50-1-Dark halo density profiles. In particular, the halos of two high-resolution UDGs are significantly underdense compared to the simulated halos, and the other low-resolution UDGs also show the same behavior to varying degrees. Thus, in addition to these inferred halo profiles being clear outliers in terms of their $R_{\rm max}$ values, our results indicate that the TNG50-1-Dark halos are also systematically overdense within the inner $2~{\rm kpc}$ compared to this sample of field UDGs. This leaves open the question of whether baryonic feedback or dark matter physics like self-interactions can alleviate the discrepancy sufficiently.

\section{Late-forming halos}
\label{sec:late}
We have seen that halos with large $R_{\rm max}$ required to host the UDGs are predicted in hierarchical structure formation theory, and their abundance is much higher than expected from a Gaussian tail. In this section, we look further into these IllustrisTNG halos. We use the TNG50-1-Dark dataset since our focus will be on the concentration, spin, and assembly history of these halos. In Fig.~\ref{fig:tng} (left), we show the number fraction of TNG50-1-Dark halos versus the normalized concentration (red), compared to a Gaussian extrapolation (dashed black). Indeed, the number of halos is {\it not} exponentially suppressed toward low $c_{200}$. Instead, it follows a power-law behavior in $|\log(c_{200}/c^{\rm median}_{200})|$ to low concentrations. For $\log(c_{200}/c^{\rm median}_{200})= -0.4$, the Gaussian extrapolation, corresponding to $\sim4\sigma$ below the median, underestimates the halo population by two orders of magnitude. The tail in $c_{200}$ values is a direct reflection for the tail in the $R_{\rm max}$ distribution shown in Fig.~\ref{fig:rmaxvmax} (left).

Before we delve further, a few comments on interpreting $c_{200}$ may be useful. When plotting the TNG50-1-Dark halos in Fig.~\ref{fig:tng} (left), we have computed their $c_{200}$ and $M_{200}$ from $V_{\rm max}$ and $R_{\rm max}$ assuming an NFW profile. Thus, $c_{200}$ and $M_{200}$ should be seen as convenient proxies for $V_{\rm max}$ and $R_{\rm max}$ from the simulations. There are other ways of computing the concentration that are profile independent, for example, $\Delta_{V/2}$, the average density within the radius where the circular velocity is $V_{\rm max}/2$ in units of the critical density~\citep{Alam:2001sn}. Given the fact that the halos with the lowest $c_{200}$ seem to have a density profile steeper than an NFW one in the inner regions, a measure like $\Delta_{V/2}$ may not identify them as being outliers. On the other hand, they are outliers because the NFW density profile is a poor fit, unlike the median halos. In this work, we have chosen to retain the simplicity of $c_{200}$ for quantitatively identifying outliers.

Fig.~\ref{fig:tng} (middle) shows the histograms of redshifts at which half of the halo mass is assembled for the TNG50-1-Dark halos with $\log_{10}(c_{200}/c^{\rm median}_{200})<-0.3$ (red) and $|\log_{10}(c_{200}/c^{\rm median}_{200})|<0.1$ (blue). We see that low-concentration (large $R_{\rm max}$) halos assemble systematically later. The formation redshift peaks around $z\sim0.5$ and $1.5$ for low and median concentration halos, respectively. Thus, the large deviations from the NFW profile apparent in density profiles of many of these halos may be tied to the late formation history of these halos and the fact that these halos may be less relaxed than median halos that formed earlier. Our fits to the circular velocity data support the hypothesis that the field UDGs may form in these halos with larger-than-expected $R_{\rm max}$ values. If so, they would also be forming later than the galaxies hosted in the median halos of the same mass. This opens up the possibility that their star formation histories (compared to more abundant galaxies) could shed light on this issue, although the small numbers of UDGs and the inherent spread in the formation times seen in Fig.~\ref{fig:tng} (middle) may make it difficult to use this as a discriminant.

The late-forming halos also have a high spin, as demonstrated in Fig.~\ref{fig:tng} (right). We calculate the spin parameter as $\lambda_{\rm halo}=J/(\sqrt{2}M_{200}V_{200}R_{200})$, where $J$ is the total angular momentum and $V_{200}=\sqrt{GM_{200}/R_{200}}$~\citep{2001ApJ...555..240B}. It is clear that the low-concentration halos have higher spin parameters on average than the median halos (see also \cite{Neto:2007vq,2007MNRAS.378...55M}. This may be relevant for creating the extended gas distribution of UDGs~\citep{2016MNRAS.459L..51A}, although AGC 114905 and AGC 242019 seem to have a normal baryonic specific angular momentum~\citep{2021A&A...651L..15M}. It would be interesting to take the late-forming IllustrisTNG halos we have identified and perform a detailed study on halo and baryonic angular momenta for the gas-rich UDGs. We note that the histograms for the spin parameters are wide with a significant overlap. To ensure that they are indeed two distinct halo populations, we conducted a Kolmogorov–Smirnov test and found that the p-value is $\sim10^{-10}$.

\begin{figure}
\centering
\includegraphics[scale=0.35]{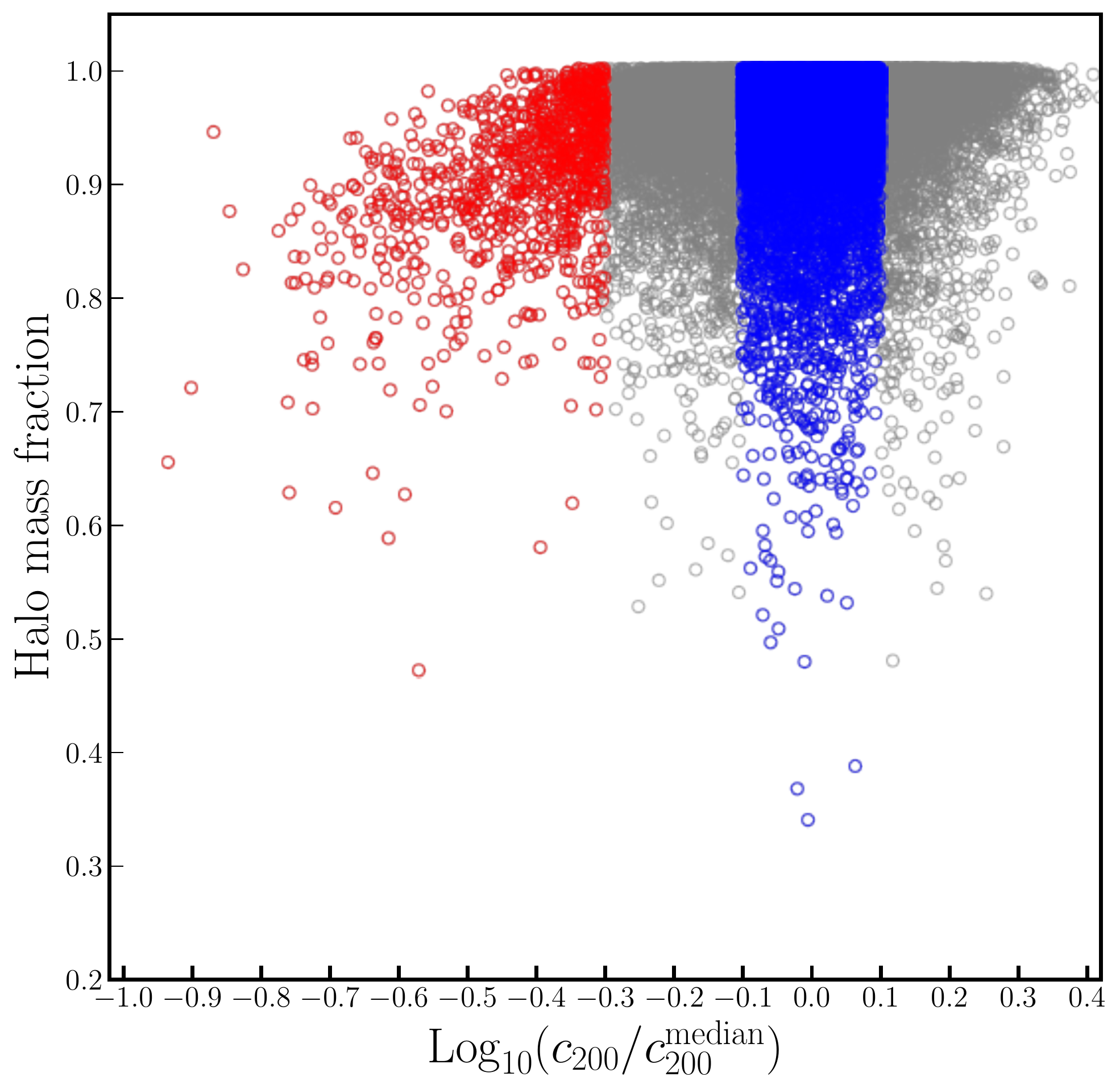}
\caption{The fraction of the dark matter mass within the virial radius that is contained in the halo (Primary Flag=1) vs normalized halo concentration for the TNG50-1-Dark halos. The red and blue colors denote
the halos satisfying $\log_{10}(c_{200}/c^{\rm median}_{200})<-0.3$ and $\left|\log_{10}(c_{200}/c^{\rm median}_{200})\right|<0.1$, respectively.}
\label{fig:fraction}
\end{figure}

For the gas-rich UDGs we consider, there is little evidence that they experienced a recent major merger. To check the merger history of the TNG50-1-Dark halos (Primary Flag=1), we obtain the fraction of the halo mass over the total group mass within the virial radius~\citep{Anbajagane:2021bvx}, as shown in Fig.~\ref{fig:fraction}. Overall, the median (blue) and low-concentration (red) halos are similar in the halo mass fraction. We note that all halos with $\log_{10}(c_{200}/c^{\rm median}_{200})\lesssim-0.5$ have a mass fraction less than $1$. Thus, there is a tendency that the extremely low concentration halos are more likely to have a relatively massive neighbor, an indication of late major mergers. However, the majority of the low-concentration halos we have identified have a mass fraction larger than $\sim0.8\textup{--}0.9$, and those halos could be ``quiet" enough to host the gas-rich UDGs discussed in this work.

To summarize this section, the IllustrisTNG halos with concentrations below about $0.3~\rm dex$ from the median are promising candidates for hosting the UDGs. Their low formation redshifts and high spins also seem to be relevant for setting the properties of the extended gas and stellar distributions of the UDGs. However, the internal densities in the simulated halos are systematically higher than what is observed in the seven field UDGs that we have analyzed.

\section{Possible solutions}
\label{sec:discussion}

\begin{figure*} [!t]
\includegraphics[scale=0.3]{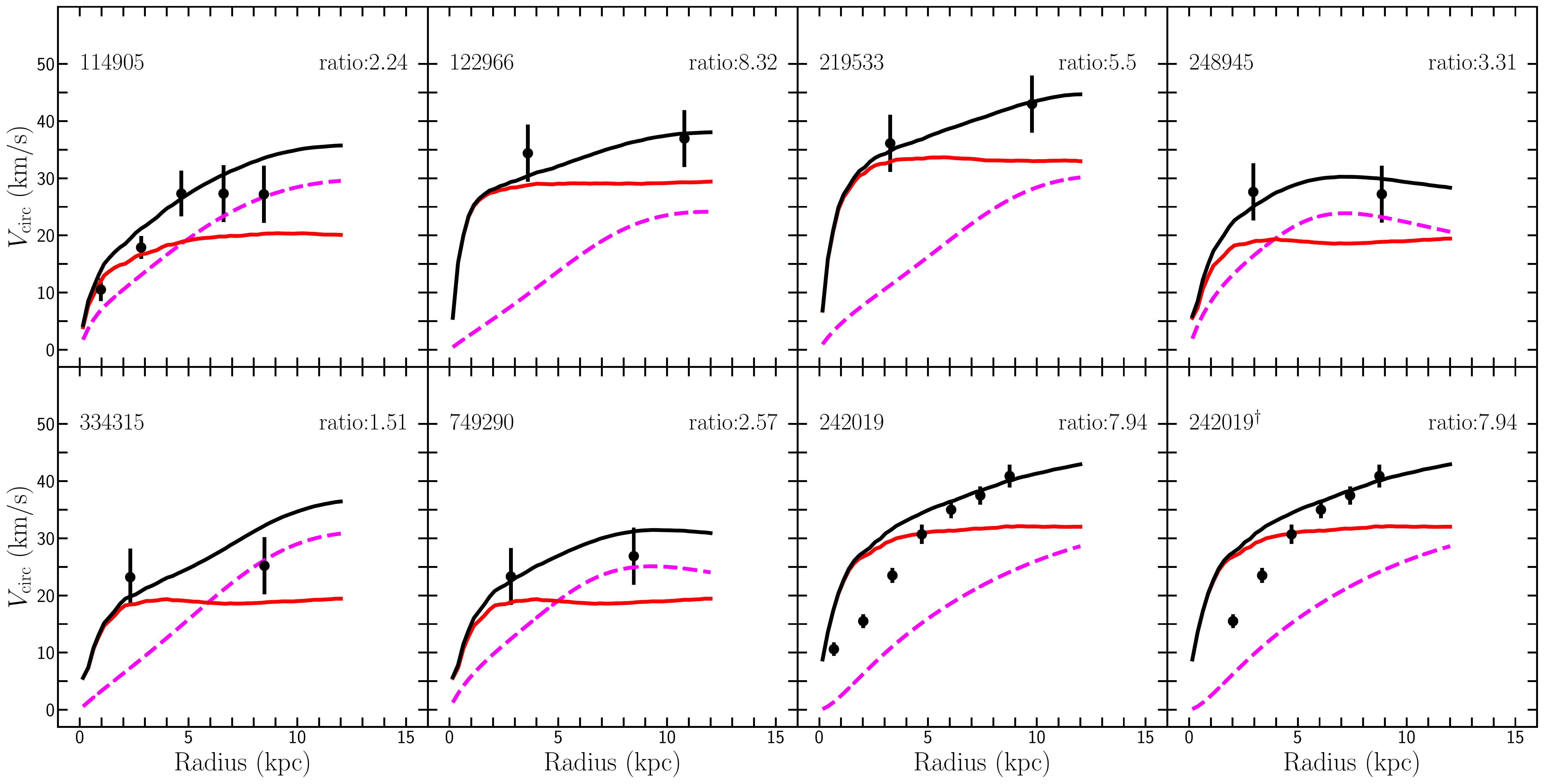}
\caption{Circular velocity profiles (solid black) constructed from example halos in TNG50-1-Dark simulations (solid red) and the baryonic contribution (dashed magenta)  compared to observational data (black circles). Each panel also shows the halo-to-baryon mass ratio in the upper right corner. All other details are the same as in Fig.~\ref{fig:fits}.}
\label{fig:fitstng}
\end{figure*}

\begin{figure*}
\centering
\includegraphics[scale=0.3]{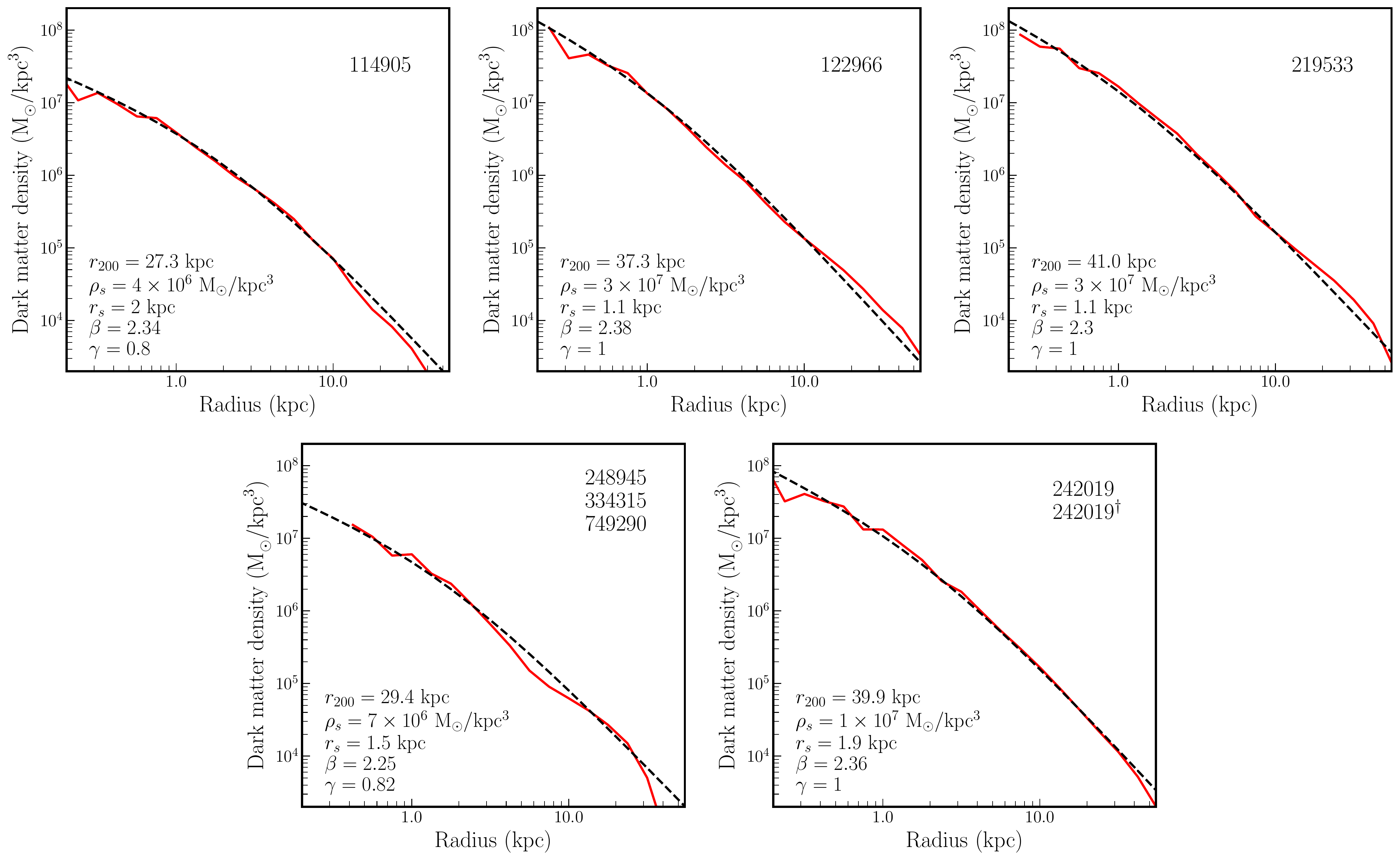}
\caption{Dark matter density profiles of the TNG50-1-Dark halos adapted in Fig.~\ref{fig:fitstng} (solid red) vs the fits using the DPL density profile in Eq.~\ref{eq:dpl} (dashed black). The parameter values of the DPL profile are shown in each panel.}
\label{fig:tngdensity}
\end{figure*}

Our analysis indicates that finding a consistent solution for the UDGs will require some modifications of the inner regions of the simulated halos. Such modifications are constrained by the fact that these galaxies are gas-rich and the ratio of baryon to halo masses is constrained in our fits to be close to the cosmological baryon fraction. Here we discuss possible ways in which the TNG50-1-Dark halos we have identified could be modified to provide a better fit to the circular velocity profiles of the field UDGs.

To further highlight the tension and modifications that are needed, we have tried to find TNG50-1-Dark halos with circular velocity profiles that are close to those of the UDGs, and the results are shown in Fig.~\ref{fig:fitstng}. Note that we do not perform a detailed fit here. Instead, we identify example TNG50-1-Dark halos that seem consistent with the observed sample in terms of the required $V_{\rm max}$. This exercise reveals critical issues exemplified by the comparisons to the two well-resolved circular velocity profiles. For AGC 242019, the ratio of simulated halo to observed baryon masses is $M_{200}/M_{\rm baryon}\approx8$, above the lower limit expected from the cosmological baryon fraction. This is not surprising, as this UDG is dark matter dominated. However, the inner density of the simulated halo is too high for AGC 242019.

AGC 114905 is even more difficult to explain, as the gas content within the inner $10~{\rm kpc}$ dominates over dark matter and stellar disk. In order to get low enough $V_{\rm max}$, we had to pick a halo with $M_{200}=1.74 M_{\rm baryon}$, a factor of $3$ below the expected halo mass from the cosmological baryon fraction, which is already very conservative. Even with such a small halo mass, the inner density needs to be further reduced. This is also clear from the inferred core size $r_c\approx6.7~{\rm kpc}$ and the inner logarithmic slope $\gamma\approx0.09$ from the Read and DPL fits, respectively. It should be noted that a low dark matter density could be in tension with disk stability requirements unless the inclination is lower or the gas dispersion is higher~\citep{2022MNRAS.512.3230M,2022arXiv220208678S}. Deeper \textsc{Hi} observations and more refined simulations are needed to further address the stability issue.

For the five low-resolution UDGs, the circular velocity profiles of the simulated halos are in good agreement with the observed ones. However, since they have only two datapoints, it is difficult to make an accurate assessment. Furthermore, for AGC 334315, AGC 749290, and AGC 248945, the simulated halo masses are too low to be consistent with the cosmological limit, as in the case of AGC 114905. It would be fascinating to see what will be revealed when higher-resolution measurement data become available for these UDGs.

Fig.~\ref{fig:fitstng} also shows that the circular velocity profiles of the simulated halos are surprisingly flat for $r\gtrsim2~{\rm kpc}$, indicating that their density profiles significantly deviate from the NFW profile. In fact, the DPL profile in Eq.~\ref{eq:dpl} can provide a good fit as shown in Fig.~\ref{fig:tngdensity}. For the sample of five simulated halos we consider, we find the transition radius $r_s\approx1\textup{--}2~{\rm kpc}$, $\gamma\sim1$, and $\beta\approx2.3$, drastically different from an NFW halo with a low concentration. The result further indicates that ``low-concentration" TNG50-1-Dark halos we have identified are not those expected from a simple extrapolation of the concentration-mass relation with an NFW profile. Thus, our application of the DPL profile to fit the UDGs is well motivated and justified. For the five halos, the fraction of the halo mass over the group mass within $r_{200}$ varies in the range of $0.8\textup{--}0.95$. Thus, we expect that they did not experience late major mergers.

\begin{figure}
\includegraphics[scale=0.35]{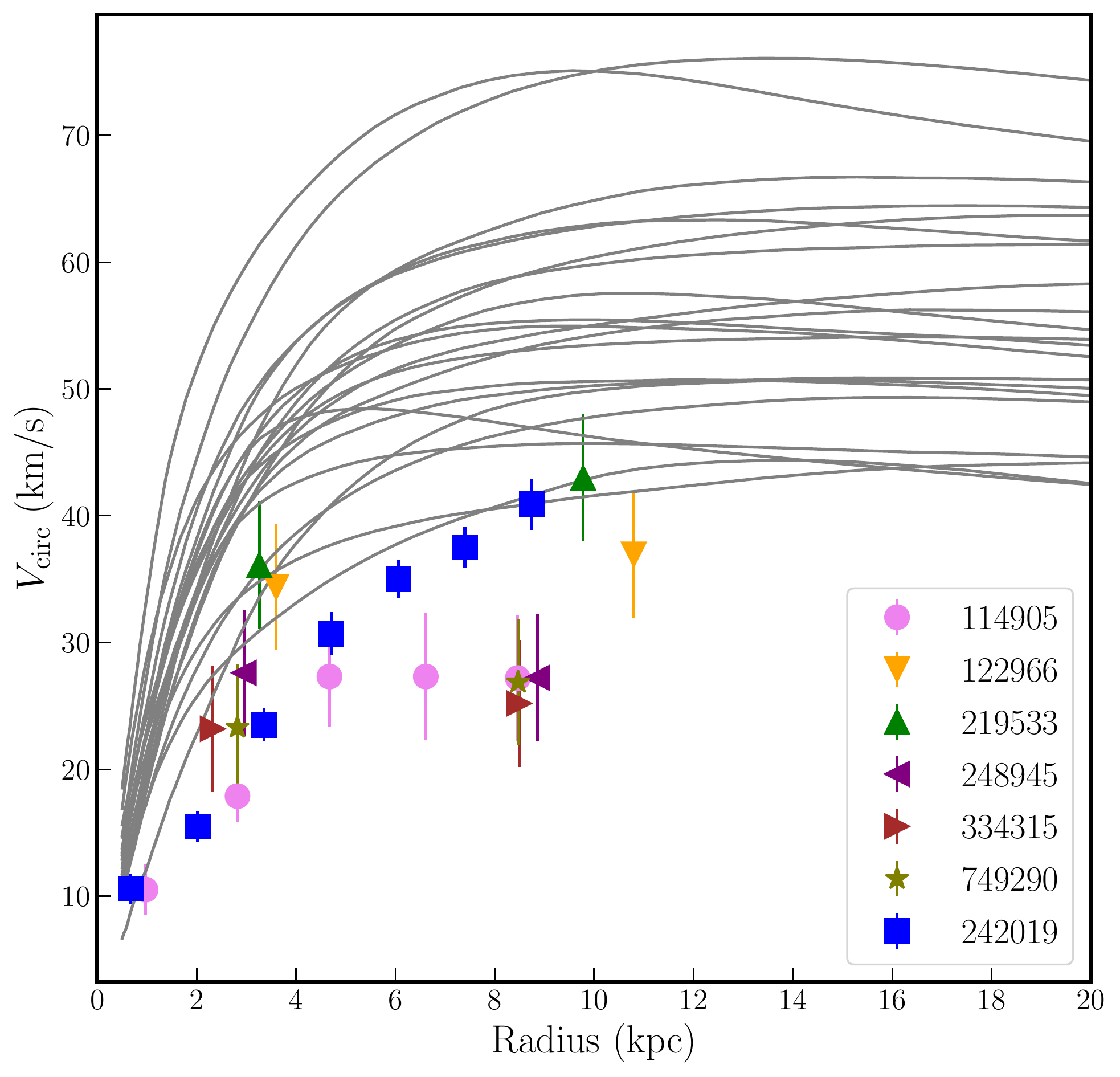}
\caption{Measured circular velocity profiles of seven isolated UDGs studied in this work (color-coded with error bars), compared to those predicted in NIHAO simulations (gray curves), based on the total mass profiles of simulated UDGs in~\cite{10.1093/mnrasl/slw210}, where $\log_{10}(M_{\rm halo}/{\rm M_\odot})=10.22\textup{--}10.85$ and $\log_{10}(M_{\rm \footnotesize \textsc{Hi}}/{\rm M_\odot})=7.22\textup{--}9.24$.}
\label{fig:nihao}
\end{figure}

\cite{10.1093/mnrasl/slw210} suggested that field UDGs could form in more normal halos using the simulations from project NIHAO (Numerical Investigation of a Hundred Astrophysical Objects); see also~\cite{2018MNRAS.478..906C}. These simulations produce extended stellar distributions, and some of the simulated galaxies are gas-rich as well. However, we find that the circular velocity profiles of the simulated UDGs in NIHAO do not match the observed UDGs in our sample as shown in  Fig.~\ref{fig:nihao}.~\footnote{Preliminary measurements in~\cite{2017ApJ...842..133L} suggested that AGC 122966, AGC 219533,  and AGC 334315 had similar rotation velocities as NIHAO UDGs, but more detailed and accurate kinematic modeling in~\cite{2019ApJ...883L..33M,2020MNRAS.495.3636M} reveals that their circular velocities are much lower.} In addition, the simulated NIHAO UDGs have overall lower gas fractions, compared to the observed gas-rich UDGs we consider.~\cite{Brook:2021veq} used a similar DPL profile to Eq.~\ref{eq:dpl} and highlighted that AGC 242019 could be hosted by a ``median" halo that has been modified by strong feedback. However, our fits show that just modifying the central density is not sufficient; large $R_{\rm max}$ is also required to get the correct $M_{\rm baryon}/M_{200}$ ratio. In fact, their best-fit model has $\beta\approx2.15$, resulting in $R_{\rm max}\approx261~{\rm kpc}$ if the density profile is extrapolated, consistent with our results.

If the gas-rich UDGs do indeed form in halos that are far from the median, then it is interesting to ask what aspects of strong feedback might change compared to the simulations based on median halos~\citep{10.1093/mnrasl/slw210,2018MNRAS.478..906C}. For example, the dynamical time $t_{\rm dyn}=2 \pi r_s/V_{\rm circ}(r_s)$ is a factor of $2\textup{--}5$ larger for $10^{10}\textup{--}10^{10.5}~{\rm M_\odot}$ halos with a concentration of $0.3\textup{--}0.6$ dex below, compared to the median halos. Thus, the low-concentration halos have a weaker potential, and it may take longer time to replenish gas necessary for bursty start formation, and the efficiency of feedback could be decreased. In addition, these low-concentration halos formed at $z \lesssim 1$ (see Fig.~\ref{fig:tng}), which reduces the time for star formation. Based on these arguments, we expect that feedback will have a lesser impact on the field UDGs compared to galaxies hosted by median concentration halos. Whether the reduced feedback efficiency is sufficient to explain the circular velocity profiles of the observed field UDGs, with gas fractions close to the cosmological value, remains to be seen.

Another possibility to consider is self-interacting dark matter (SIDM; see~\cite{2018PhR...730....1T} for a review). The self-interactions thermalize the inner halo and heat up dark matter particles, thereby reducing the inner halo density. Studies have shown that SIDM can explain the diverse rotation curves over a wide mass range of spiral galaxies~\citep{2019PhRvX...9c1020R,2020JCAP...06..027K} and the properties of UDGs in clusters and groups~\citep{2019MNRAS.485..382C,2020PhRvL.125k1105Y}. For field UDGs, self-interactions would need to create constant-density core sizes of few kiloparsecs to be able to lower the circular velocity at $\sim 2~{\rm kpc}$; see Fig.~\ref{fig:rmaxvmax} (right).

The low ``concentrations'' inferred in the UDGs imply that the radius out to which their host halos are thermalized~\citep{Kaplinghat:2015aga} would be a small fraction of $r_s$ for self-interacting cross sections over mass of order $1~\rm cm^2 ~g^{-1}$, which have been used to explain the rotation curves of spiral galaxies~\citep{2019PhRvX...9c1020R}. Conversely, we expect core size constraints in the UDGs to provide significant constraints on SIDM models. In this regard, the two high-resolution UDGs AGC 114905 and AGC 242019 present different challenges. AGC 114905 is completely dominated by gas, which will need to be folded in to determine the SIDM density profile~\citep{Kaplinghat:2013xca}, while the core size inference in AGC 242019 is highly dependent on the innermost data point. More well-resolved circular velocity curves for the UDGs are required to make further progress on this front.

An interesting connection is that some of the field UDGs are close to the higher end of the velocity range relevant for the Milky Way dwarf spheroidal satellites, approximately $5\textup{--}30~\rm km ~s^{-1}$. Recent studies have argued for cross sections much larger than $1~\rm cm^2 ~g^{-1}$ at those velocities based on the observed diversity in the central dark matter densities of the satellites~\citep{Nishikawa:2019lsc,Zavala:2019sjk,Kahlhoefer:2019oyt,Sameie:2019zfo,Correa2021,Jiang:2021foz}. Thus, in conjunction with the Milky Way satellites, we expect the field gas-rich UDGs to be a strong test of SIDM models.

\section{Conclusions}
\label{sec:conclusions}

We have analyzed the circular velocity profiles of seven gas-rich UDGs in the field and inferred properties of their host dark matter halos. The preferred halo mass is $M_{200}\sim10^{10}\textup{--}10^{11}~{\rm M_\odot}$ ($V_{\rm max}\sim30\textup{--}50~{\rm km ~s^{-1}}$), and the resulting ratio of baryon-to-halo masses is $\sim 0.07$. The baryons in these UDGs are mostly in the form of \textsc{Hi} gas, with the ratio of gas to stellar masses in the range of $3\textup{--}50$.

We have argued in this paper that simulated halos with $c_{200}$ values about $0.3$ dex or lower than the median seem to be the best possible hosts for the field UDGs we have analyzed. Remarkably, we find that such low-concentration (large $R_{\rm max}$) halos are formed in cold dark matter models (and by extension other models with a similar initial power spectrum).  We have demonstrated this using the IllustrisTNG simulations, in particular the dark-matter-only run, and showed that the distribution of halo concentrations has a large non-Gaussian tail to low concentrations. We have identified that this population of simulated hosts, on average, forms later than the median halos and has higher spins. It is possible that these features may be tied to the unusual properties of the gas-rich UDGs.

The gas-rich field UDGs we have analyzed are outliers in terms of their inferred halo properties. We emphasize that the host halos are low-concentration in the sense of having large $R_{\rm max}$, but their density profiles could depart drastically from the NFW profile in both inner and outer regions. The low dark matter densities and the odd halo properties of these gas-rich field UDGs provide an exciting opportunity to test both feedback models and dark matter models. To take advantage of this opportunity, we need to discover more field UDGs and obtain well-resolved circular velocity profiles for them.

\section*{Acknowledgements}

We thank Volker Springel and collaborators for the IllustrisTNG simulation data and the user-friendly data access interface, and Salvador Cardona Barrero for clarifications on the NIHAO data. M.K. is supported by the NSF under award No. 1915005. H.-B.Y. is supported by the U.S. Department of Energy under grant No. de-sc0008541, the John Templeton Foundation under grant ID \#61884, and NASA under grant No. 80NSSC20K0566. F.F. and P.E.M.P. are supported by the Netherlands Research School for Astronomy (Nederlandse Onderzoekschool voor Astronomie, NOVA), Network 1, Project 10.1.5.6.  The opinions expressed in this publication are those of the authors and do not necessarily reflect the views of the John Templeton Foundation.

\appendix

\begin{table*}

\centering

\begin{tabular}{ c | c | c | c | c }
\hline
{\rm AGC ID} &  $V_{\rm sys}$ (km/s) & $D$ (Mpc) & $i$ (deg) & References \\
\hline
114905 & $5435$ & $67 \pm 1$ & $26.1^{+0.2}_{-0.1}$ & \cite{2017ApJ...842..133L, 2020MNRAS.495.3636M, 2022MNRAS.512.3230M} \\
122966 & $6509$ & $90 \pm 5$ & $34 \pm 5$ & \cite{2017ApJ...842..133L, 2020MNRAS.495.3636M} \\
219533 & $6384$ & $96 \pm 5$ & $42 \pm 5$ & \cite{2017ApJ...842..133L, 2020MNRAS.495.3636M} \\
248945 & $5703$ & $84 \pm 5$ & $66 \pm 5$ & \cite{2017ApJ...842..133L, 2020MNRAS.495.3636M} \\
334315 & $5107$ & $73 \pm 5$ & $45 \pm 5$ & \cite{2017ApJ...842..133L, 2020MNRAS.495.3636M} \\
749290 & $6516$ & $97 \pm 5$ & $39 \pm 5$ & \cite{2017ApJ...842..133L, 2020MNRAS.495.3636M} \\
242019 & $1840.4$ & $30.8 \pm 1.5$ & $73.0\textup{--}70.2$ &  \cite{2021ApJ...909...20S} \\
\hline
\end{tabular}

\caption{ Additional information about properties of the gas-rich UDGs We Consider. From left to right columns: galaxy AGC ID, systemic velocity, distance, inclination, source of the data. For AGC 114905,~\cite{2022MNRAS.512.3230M} reports another set of distance and inclination values $D=76 \pm 5~{\rm Mpc}$ and $i=32 \pm 3^{\rm o}$. In our fits, we take the values listed in the table for AGC 114905, see~\cite{2022MNRAS.512.3230M} for details about measuring the inclination. For AGC 242019, \cite{2021ApJ...909...20S} reports that the inclination varies from $i=73.0^{\rm o}$ to $70.2^{\rm o}$ for the distance in the range $0.67\textup{--}8.75~{\rm kpc}$.}

\label{table:parameters3}
\end{table*}

\begin{figure*} [t!]
\centering
\includegraphics[scale=0.3]{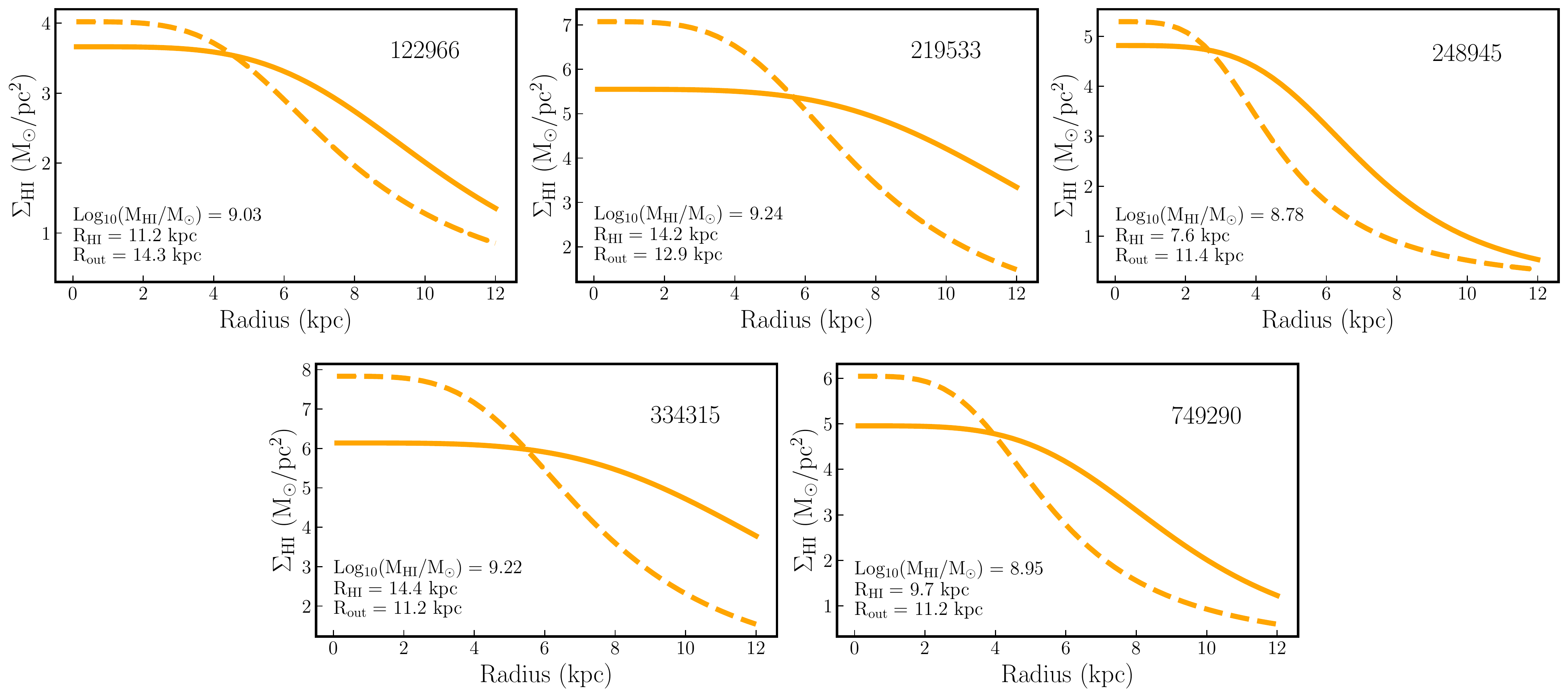}
\caption{\textsc{Hi} gas surface mass density profiles of the low-resolution UDGs for two sets of model parameters $n=4$ (solid) and $2.5$ (dashed), where $m=4$ in both cases; see Eq.~\ref{eq:gas}. The central surface density $\Sigma_{\rm \footnotesize \textsc{Hi}}(0)$ is set by demanding the predicted \textsc{Hi} mass within $R_{\rm out}$ to be the median value of \textsc{Hi} mass in~\cite{2021ApJ...909...19G}, as denoted in each panel. Our MCMC routine assumes a flat prior for $M_{\rm \footnotesize \textsc{Hi}}$, and thus $\Sigma_{\rm \footnotesize \textsc{Hi}}(0)$ varies accordingly.}
\label{fig:gas2}
\end{figure*}

\begin{figure*}
\centering
\includegraphics[scale=0.4]{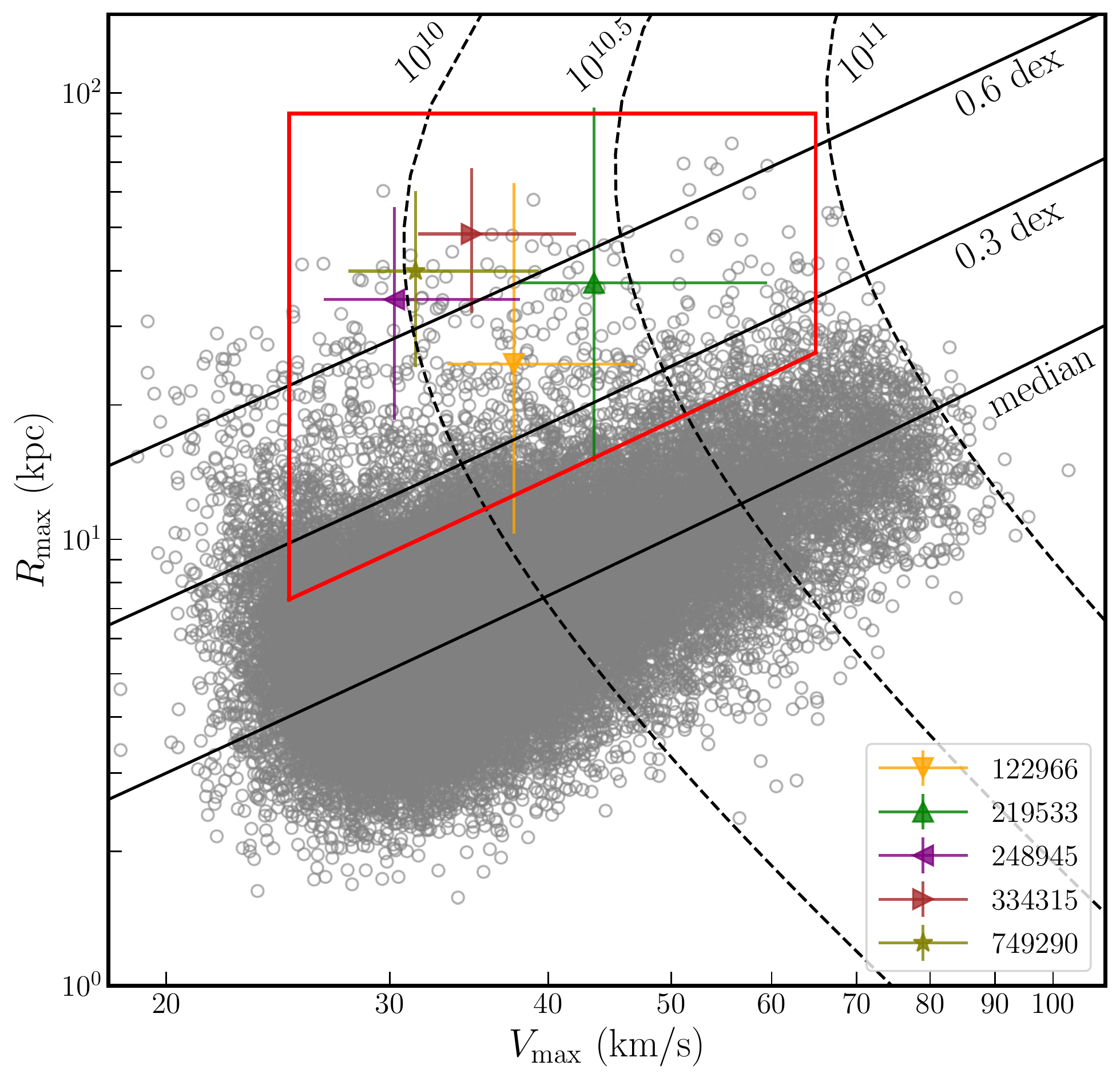}
\caption{Similar to the left panel of Fig.~\ref{fig:rmaxvmax} for the low-resolution UDGs, but with a gas surface density profile of $n=2.5$ and $m=4$ (dashed error bars). Compared to the case with $n=4$ and $m=4$ in Fig.~\ref{fig:rmaxvmax}, the inferred $R_{\rm max}$ values are somewhat smaller. }
\label{fig:rmaxvmaxn25}
\end{figure*}

\begin{figure*}
\includegraphics[scale=0.3]{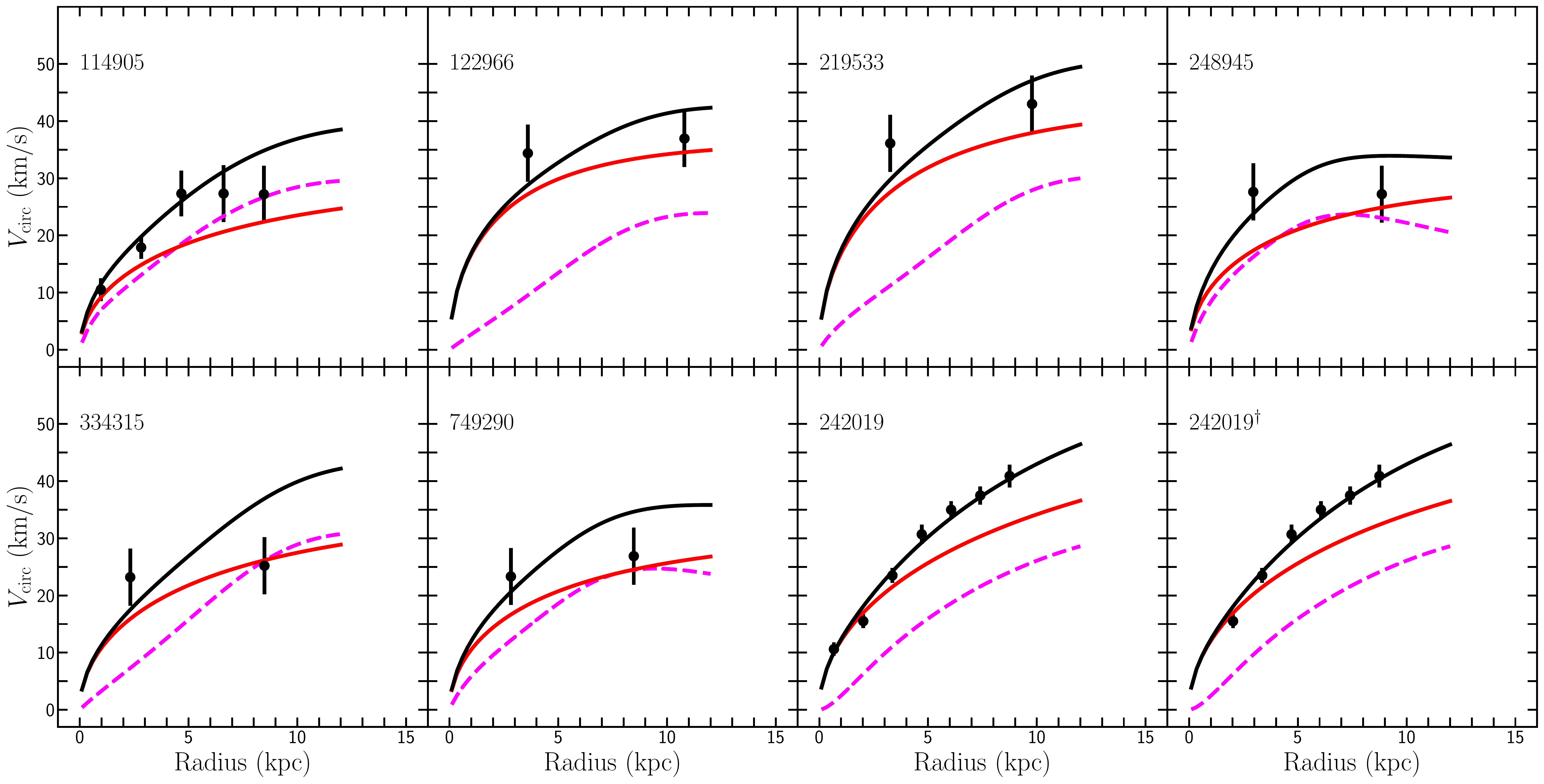}
\caption{Circular velocity profiles of the best fit model (solid black), including halo (solid red) and baryonic (dashed magenta) contributions compared to observational data (black dots), where we assume the NFW density profile. For AGC 242019$^{\dagger}$, the fit is performed without including the innermost datapoint. The observational data are taken from~\cite{2022MNRAS.512.3230M,2020MNRAS.495.3636M,2021ApJ...909...20S}.
}
\label{fig:fitsnfw}
\end{figure*}

\begin{figure*}
\centering
\includegraphics[scale=0.4]{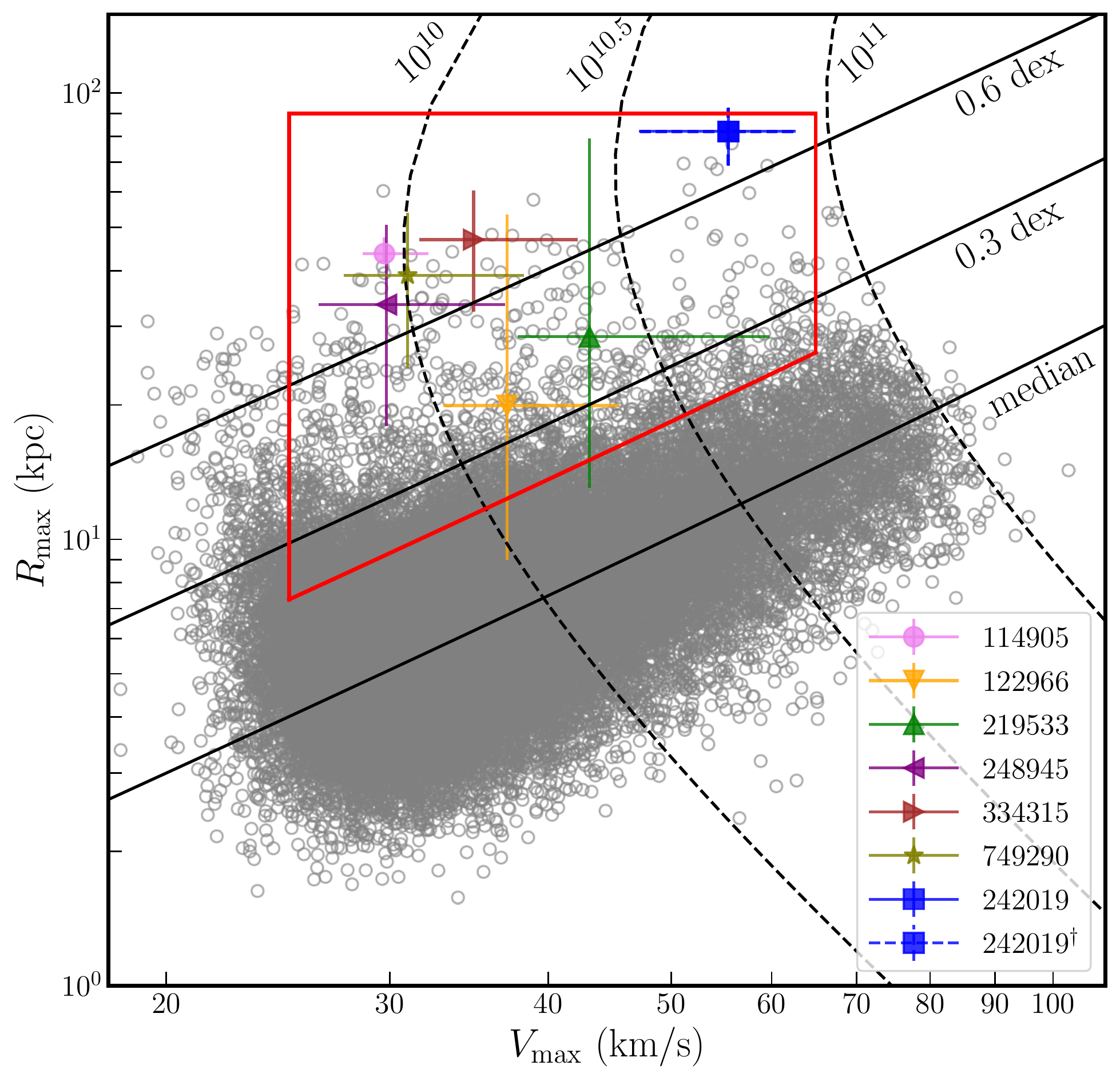}
\includegraphics[scale=0.4]{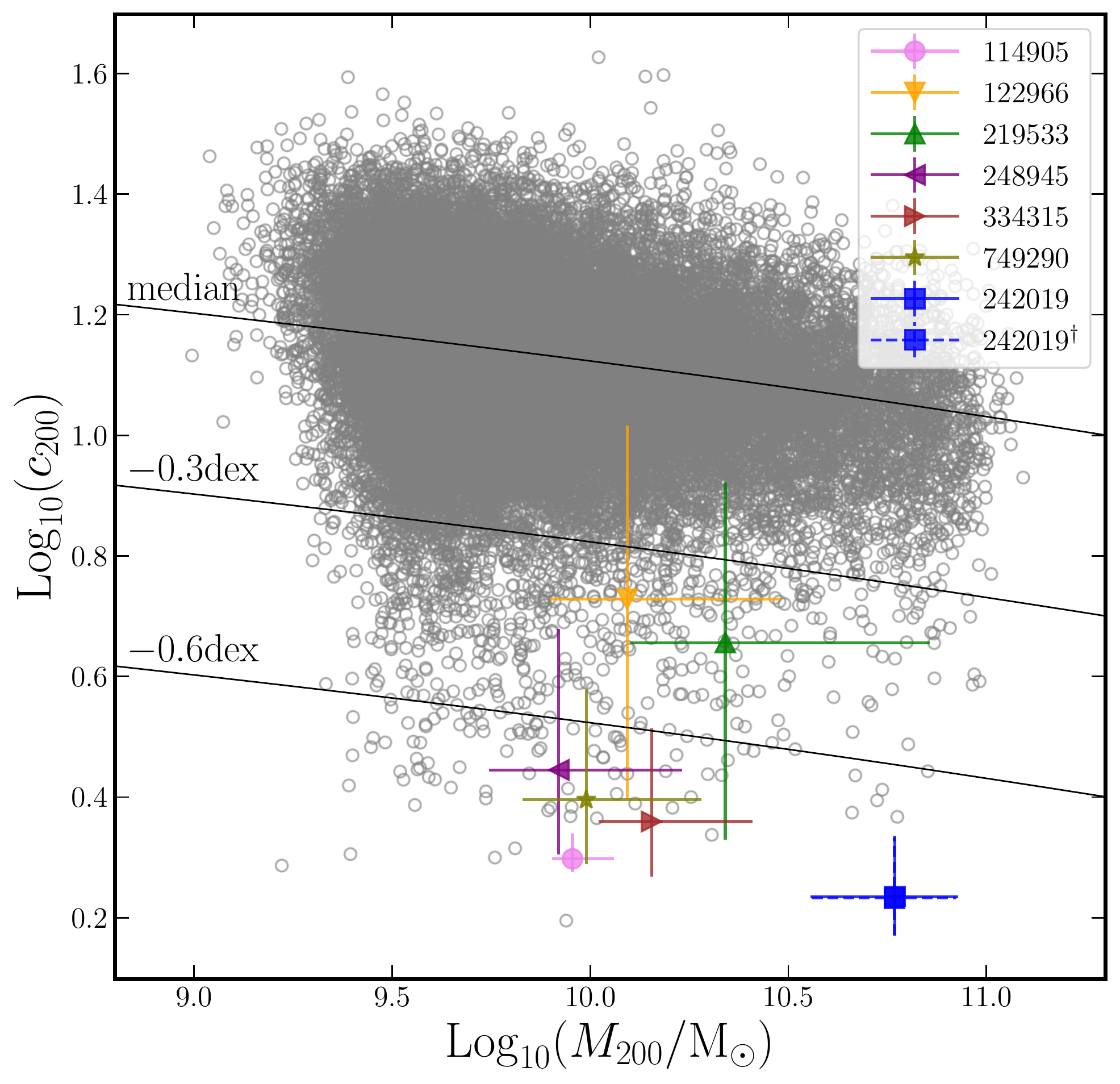}
\caption{{\it Left:} similar to the left panel of Fig.~\ref{fig:rmaxvmax}, but for fits with the NFW density profile. The inferred $V_{\rm max}$ and $R_{\rm max}$ values are similar to those from the Read fits. For AGC 114905 and AGC 242019$^{\dagger}$ (overlaps with AGC 242019 in the $R_{\rm max}\textup{--}V_{\rm max}$ plane), the difference is relatively large, because the halo is not cored in this case. {\it Right}: $c_{200}\textup{--}M_{200}$ distributions from the NFW fits, complementary to the $R_{\rm max}\textup{--}V_{\rm max}$ distributions shown in the left panel.}
\label{fig:rmaxvmaxNFW}
\end{figure*}

\begin{figure*}
\includegraphics[scale=0.3]{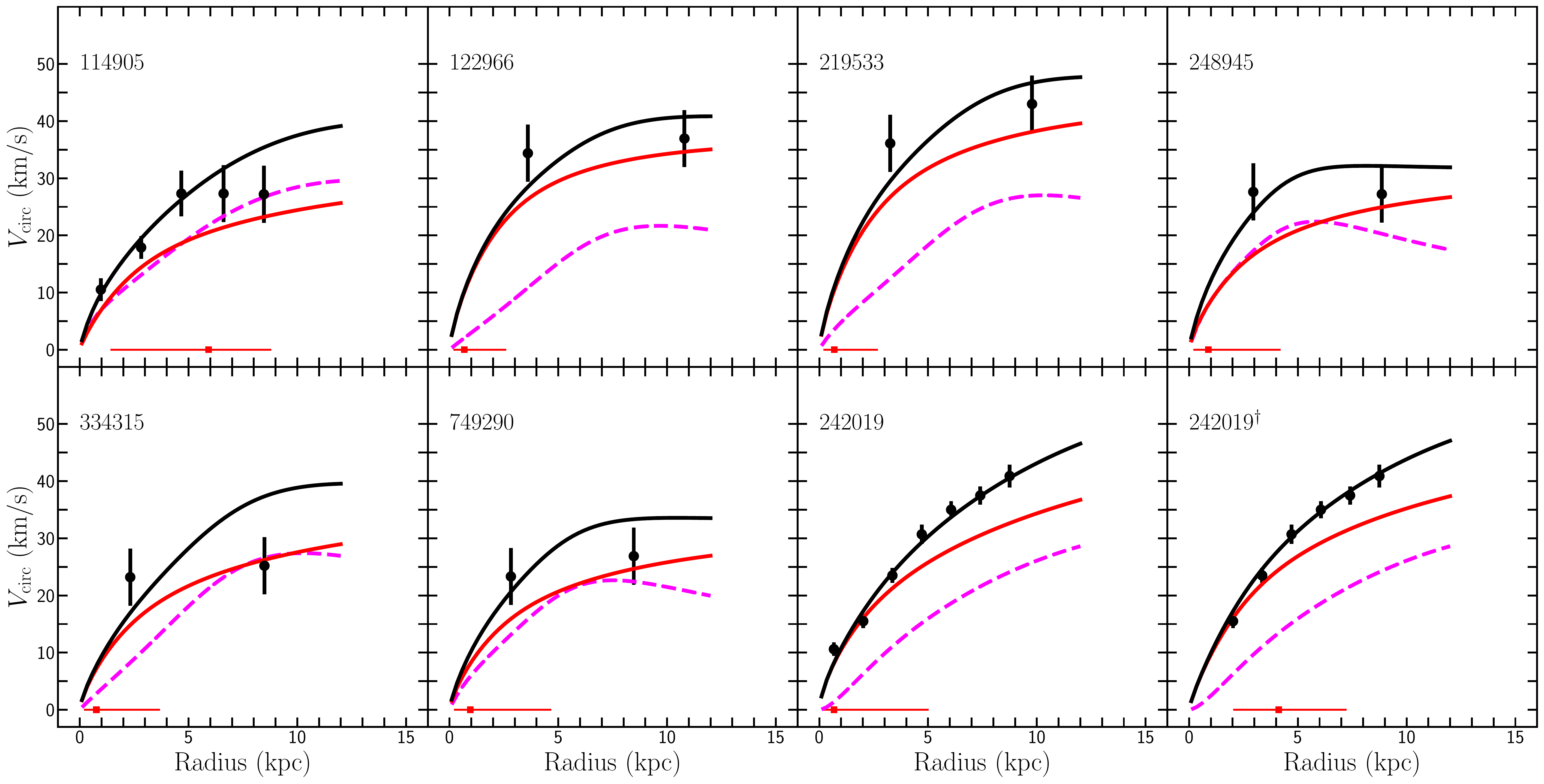}
\caption{Circular velocity profiles of the best fit model (solid black), including halo (solid red) and baryonic (dashed magenta) contributions compared to observational data (black circles), where we assume the general Read profile. For AGC 242019$^{\dagger}$, the fit is performed without including the innermost datapoint. The observational data are taken from~\cite{2022MNRAS.512.3230M,2020MNRAS.495.3636M} and ~\cite{2021ApJ...909...20S}.
}
\label{fig:fitsreadgeneral}
\end{figure*}

\begin{figure*}
\centering
\includegraphics[scale=0.3]{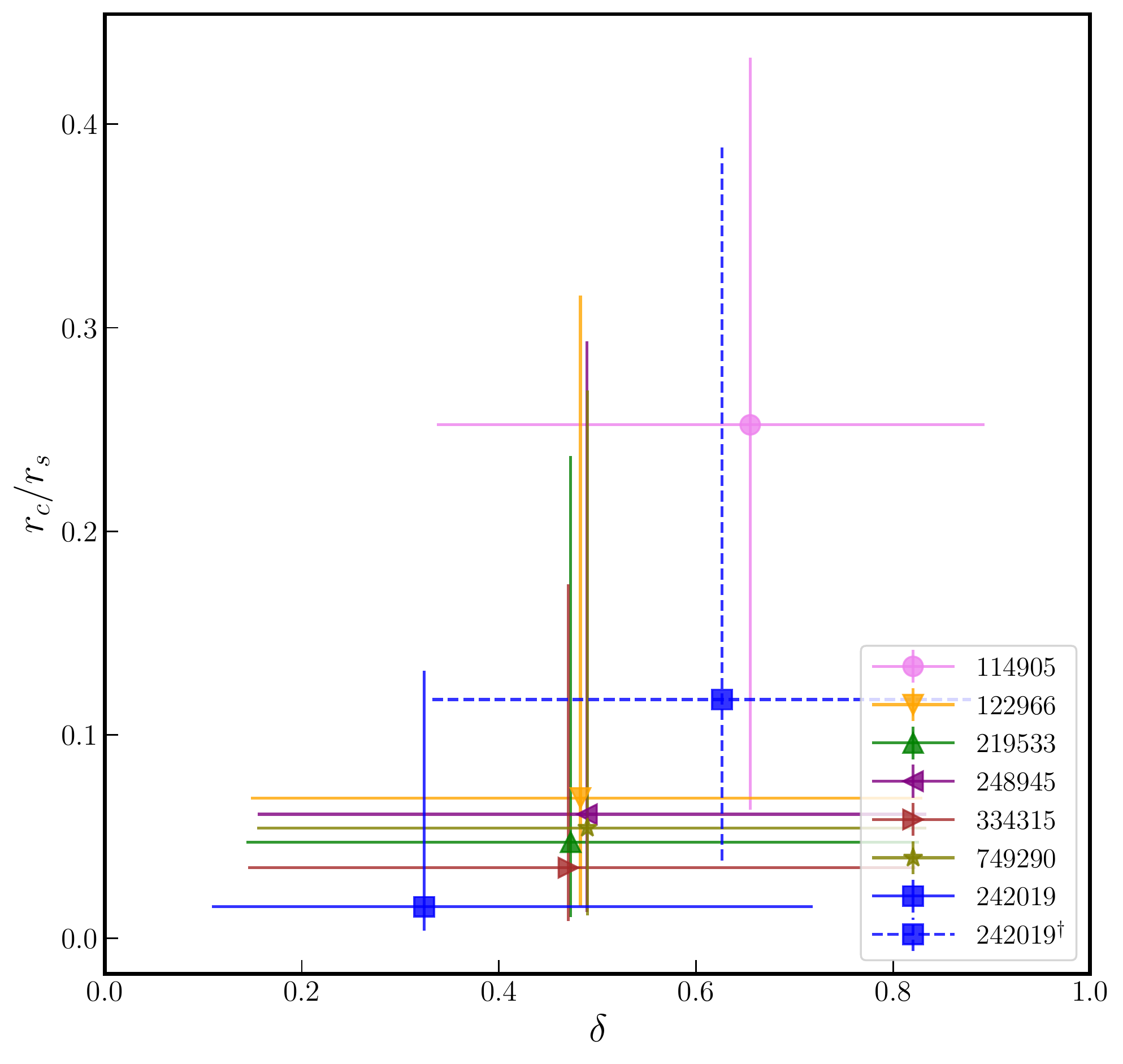}~~~
\includegraphics[scale=0.3]{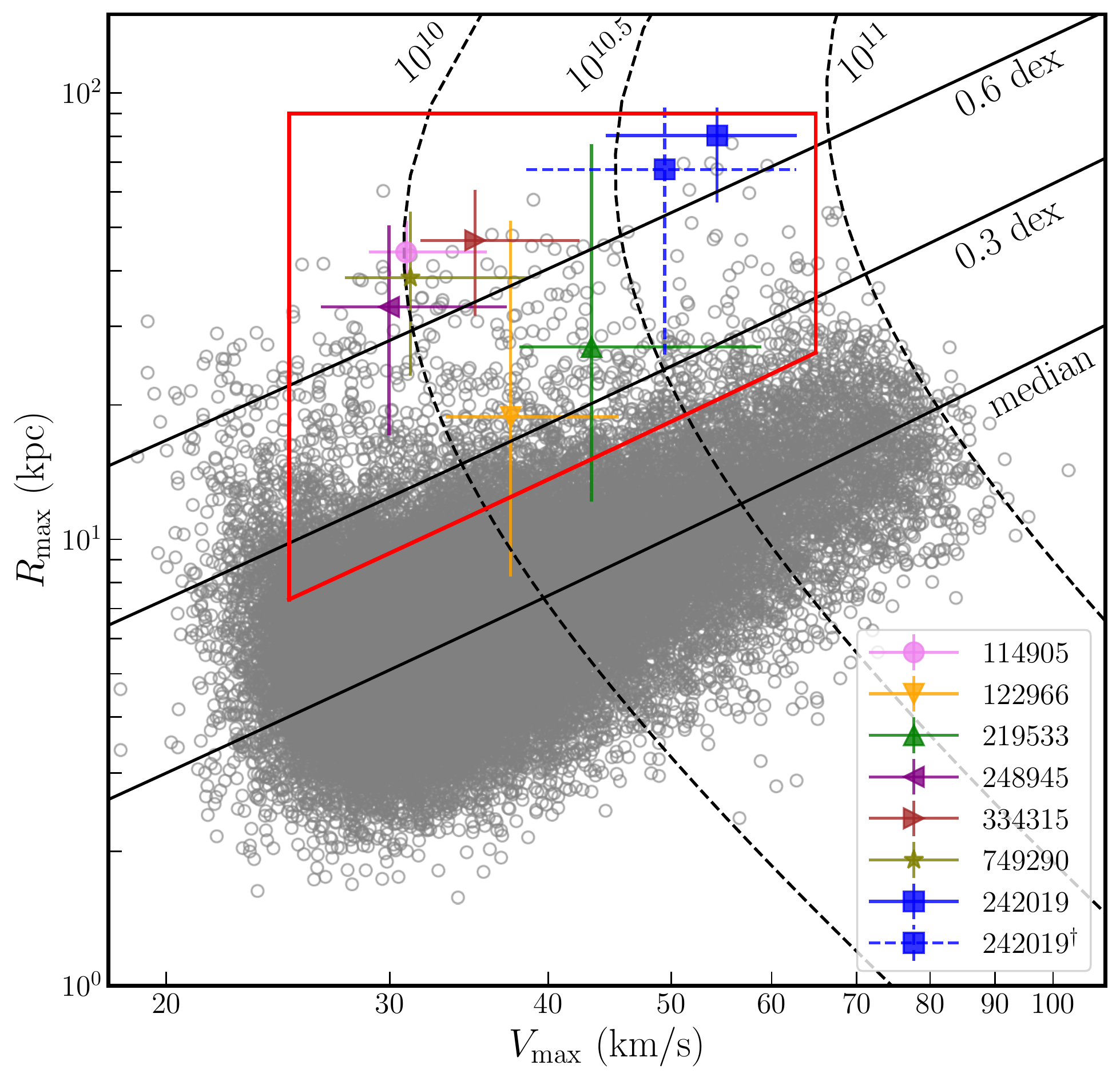}
\caption{{\it Left:} inferred $r_c/r_s$ and $\delta$ distributions from the fits with the general Read profile. {\it Right:} similar to the left panel of Fig.~\ref{fig:rmaxvmax}, but with the general Read profile.}
\label{fig:rcdelta}
\end{figure*}

\begin{figure*}
\includegraphics[scale=0.3]{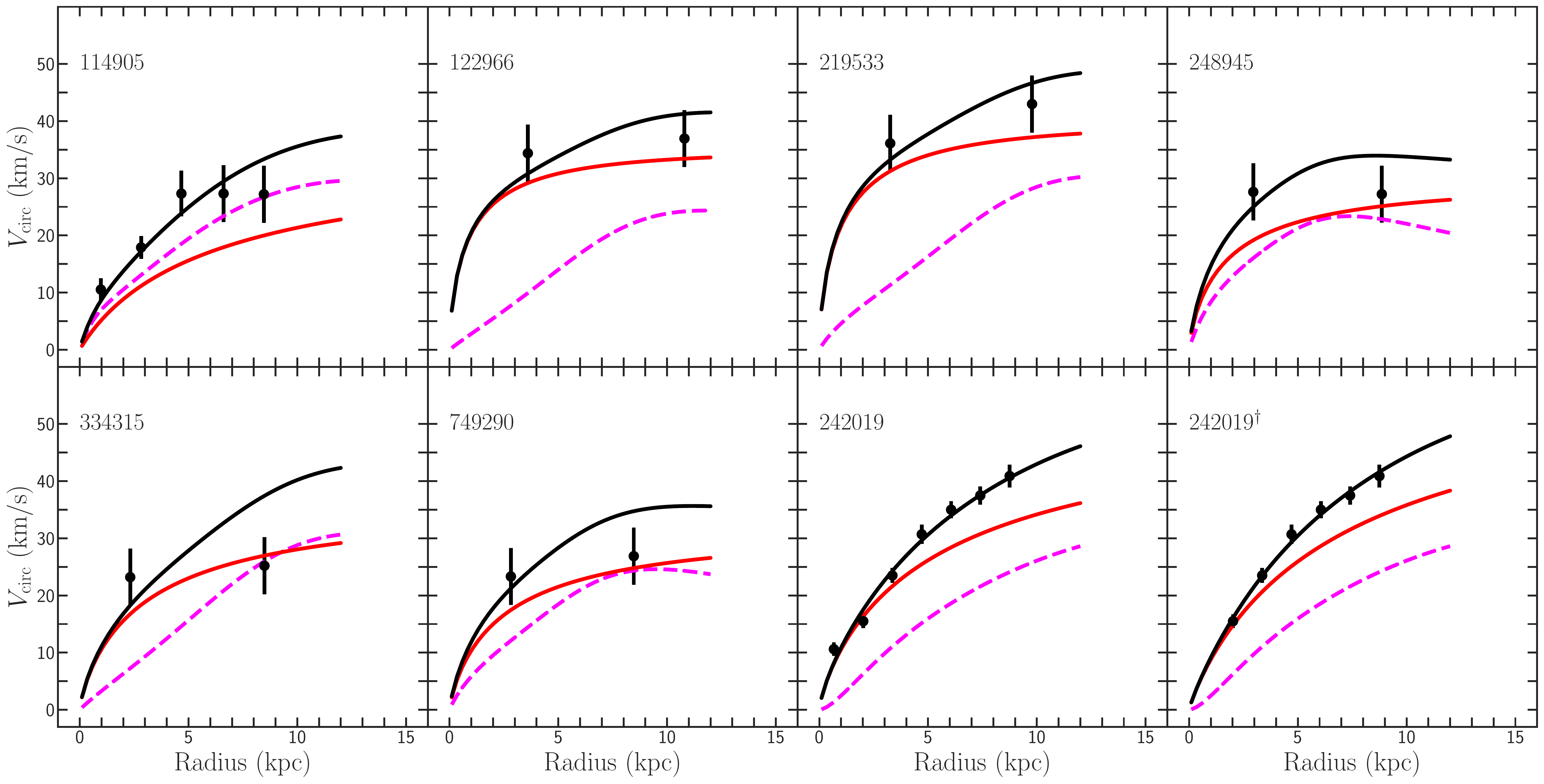}
\caption{Circular velocity profiles of the best-fit model (solid black), including halo (solid red) and baryonic (dashed magenta) contributions compared to observational data (black circles). The halo profile is the DPL profile in Eq.~\ref{eq:dpl}. For AGC 242019$^{\dagger}$, the fit is performed without including the innermost datapoint. The observational data are taken from~\cite{2022MNRAS.512.3230M,2020MNRAS.495.3636M} and ~\cite{2021ApJ...909...20S}. For the inferred parameter errors, see Table~\ref{table:parameters2}.
}
\label{fig:fits2}
\end{figure*}

\begin{figure*}
\includegraphics[scale=0.5]{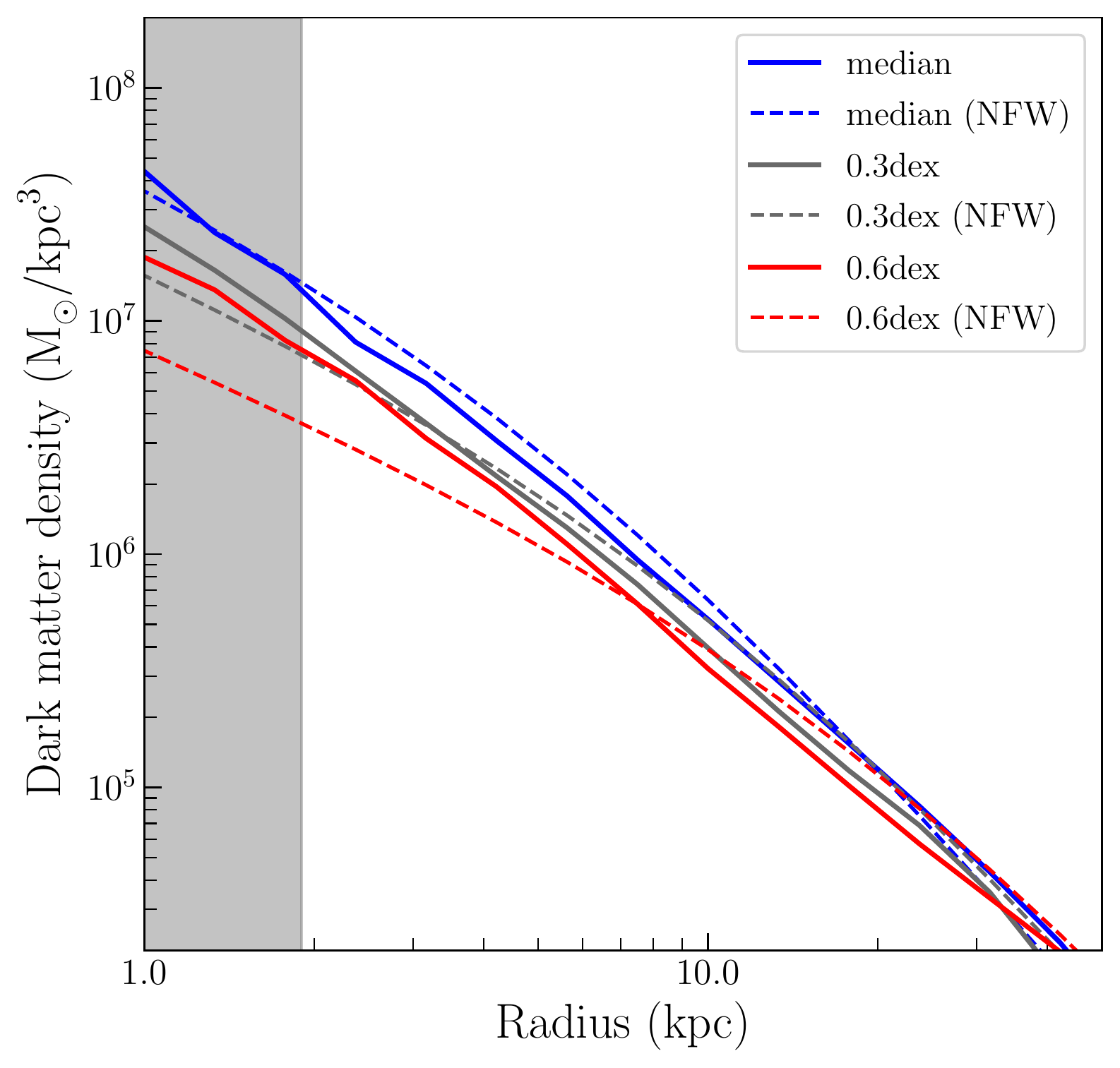}
\includegraphics[scale=0.5]{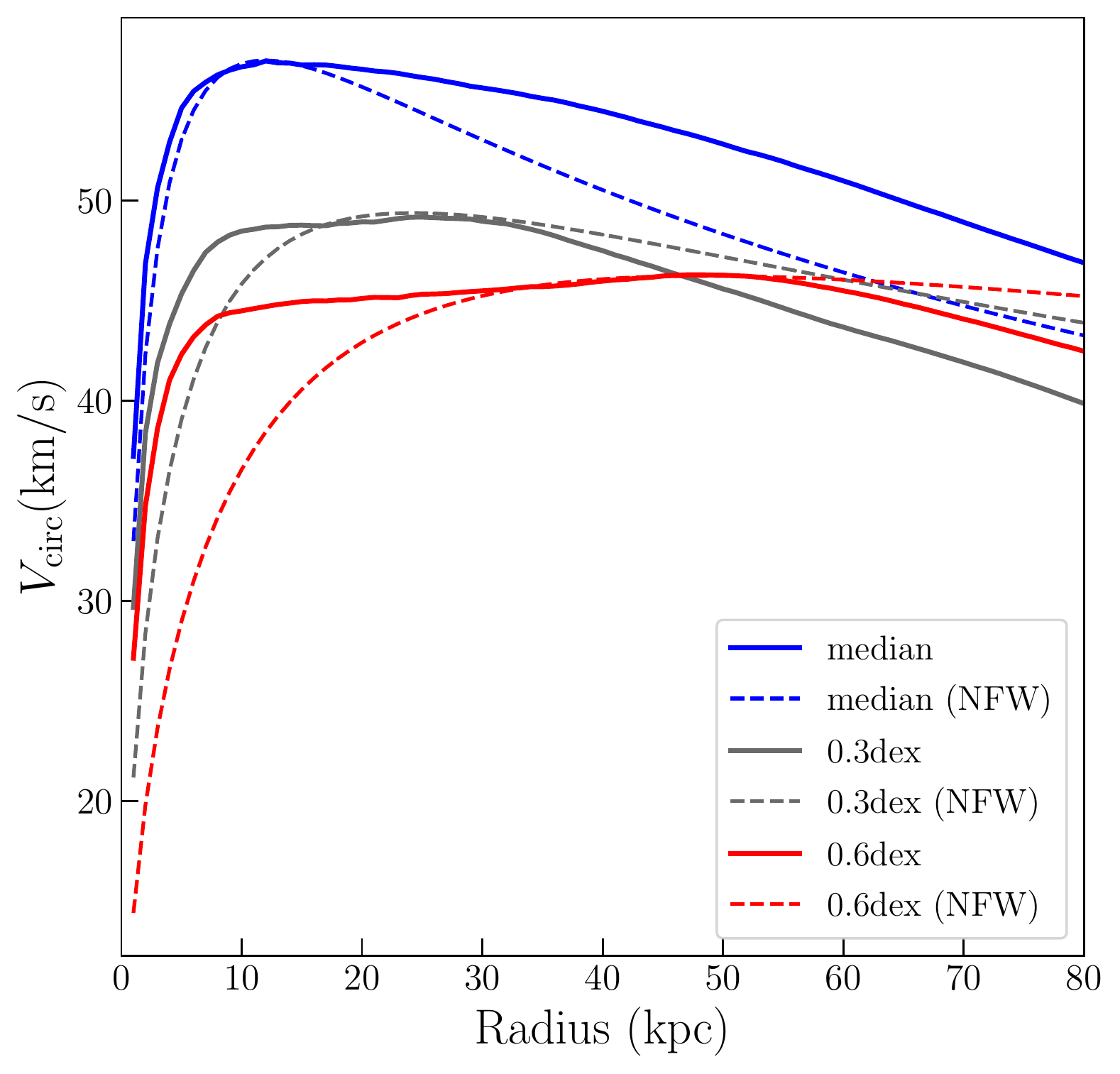}
\caption{{\it Left:} dark matter density profiles of example $M_{200}\approx3.2\times10^{10}~{\rm M_\odot}$ TNG50-1-Dark halos with a median concentration (blue) and $0.3$ dex (gray) and $0.6$ dex (red) below. The solid curves are computed from the particle data, while the dashed curves are NFW profiles using the catalog halo parameters $(V_{\rm max},~R_{\rm max})$:  $(56.9~{\rm km ~s^{-1}},~11.9~{\rm kpc})$ (blue), $(49.1~{\rm km ~s^{-1}},~24.3~{\rm kpc})$ (gray), and $(46.2~{\rm km ~s^{-1}},~48.6~{\rm kpc})$ (red). As expected, the density profile of the median halo is NFW-like, but it becomes steeper for the low concentration halos. The halo with the lowest concentration (red) has a logarithmic slope of $-1.4$ within $\sim8~{\rm kpc}$, which is approximately $R_{\rm \footnotesize \textsc{Hi}}$ of the UDGs. The shaded region indicates the resolution limit. {\it Right:} the corresponding circular velocity profiles for the halos shown in the left panel computed from their simulated mass distribution (solid), compared to the calculated ones assuming an NFW profile (dashed).}
\label{fig:halo}
\end{figure*}

In this appendix, we provide additional information about properties of the gas-rich UDGs, the \textsc{Hi} gas surface density for the low-resolution UDGs, the fits using the NFW, general Read and DPL profiles, inferred $R_{\rm max}\textup{--}V_{\rm max}$ distributions using a different \textsc{Hi} gas surface density profile, and dark matter distributions of three representative IllustrisTNG halos.

Table ~\ref{table:parameters3} provides addition information about the gas-rich UDGs considered in this work, including their systemic velocity, distance, and inclination.

Fig.~\ref{fig:gas2} shows \textsc{Hi} gas surface mass density profiles of the low-resolution UDGs for two sets of model parameters $n=4$ and $2.5$. We fix $m=4$ in both cases.

Fig.~\ref{fig:rmaxvmaxn25} shows the inferred $R_{\rm max}\textup{--}V_{\rm max}$ distributions for the low-resolution UDGs, similar to the left panel of Fig.~\ref{fig:rmaxvmax}, but using a gas surface density profile with $n = 2.5$ and $m = 4$. See Fig.~\ref{fig:gas2} for the difference in the gas surface density for $n=4$ and $2.5$.

Fig.~\ref{fig:fitsnfw} shows circular velocity profiles of the best fit model for the NFW profile. The inferred $R_{\rm max}\textup{--}V_{\rm max}$ and $c_{200}\textup{--}M_{200}$ distributions are shown in Fig.~\ref{fig:rmaxvmaxNFW}.

Fig.~\ref{fig:fitsreadgeneral} shows circular velocity profiles of the best fit model for the general Read profile. The inferred $r_c/r_s$ and $\delta$ distributions, as well as the $R_{\rm max}\textup{--}V_{\rm max}$ distributions, are shown in the left and right panels of Fig.~\ref{fig:rcdelta}, respectively.

Fig.~\ref{fig:fits2} shows circular velocity profiles of the best-fit model for the DPL profile in Eq.~\ref{eq:dpl}. The model parameters are summarized in Table~\ref{table:parameters2}, and the inferred $R_{\rm max}\textup{--}V_{\rm max}$ distributions are shown in Fig.~\ref{fig:rmaxvmax2}.

Fig.~\ref{fig:halo} shows dark matter density profiles of three TNG50-1-Dark halos with three different concentrations (left panel) and their corresponding profiles of circular velocities (right panel).

 \clearpage

\bibliography{example}{}
\bibliographystyle{aasjournal}

\end{document}